\documentclass[prd,reprint,twocolumn,showpacs,nofootinbib]{revtex4}

\usepackage{amsfonts}
\usepackage{amsmath}
\usepackage{amssymb}
\usepackage{aas_macros}
\usepackage{upgreek}
\usepackage{bm}
\usepackage{dcolumn}
\usepackage{epsfig}
\usepackage{graphicx}
\usepackage{graphics}
\usepackage{placeins}
\usepackage{subfigure}
\usepackage{dsfont}
\usepackage{subfigure}
\newcommand{\mb}[1]{\mathbf{#1}}
\newcommand{\unit}{\mathds{1}}
\newcommand{\zero}{\mb{0}}

\usepackage{color}

\begin{document}

\title{When is a gravitational-wave signal stochastic?}

\author{Neil J.~Cornish}
\affiliation{Department of Physics, Montana State University, 
Bozeman, MT 59718, USA.}
\author{Joseph D.~Romano}
\affiliation{Department of Physics and Astronomy 
and Center for Gravitational-Wave Astronomy,
University of Texas at Brownsville, Brownsville, TX 78520, USA.}

\date{\today}

%%%%%%%%%%%%%%%%%%%%%%%%%%%%%%%%%%%%%%%%%%%%%%%
\begin{abstract}

We discuss the detection of gravitational-wave backgrounds in the
context of Bayesian inference and suggest a practical definition of
what it means for a signal to be considered stochastic---namely, that
the Bayesian evidence favors a stochastic signal model over a
deterministic signal model.  A signal can further be classified as
Gaussian-stochastic if a Gaussian signal model is favored.  In our
analysis we use Bayesian model selection to choose between several
signal and noise models for simulated data consisting of uncorrelated
Gaussian detector noise plus a superposition of sinusoidal signals
from an astrophysical population of gravitational-wave sources.  For
simplicity, we consider co-located and co-aligned detectors with white
detector noise, but the method can be extended to more realistic
detector configurations and power spectra.  The general trend we
observe is that a deterministic model is favored for small source
numbers, a non-Gaussian stochastic model is preferred for intermediate
source numbers, and a Gaussian stochastic model is preferred for large
source numbers. However, there is very large variation between
individual signal realizations, leading to fuzzy boundaries between
the three regimes. We find that a hybrid, trans-dimensional model
comprised of a deterministic signal model for individual bright
sources and a Gaussian-stochastic signal model for the remaining
confusion background outperforms all other models in most instances.

\end{abstract}

%%%%%%%%%%%%%%%%%%%%%%%%%%%%%%%%%%%%%%%%%%%%%%%

\maketitle

%%%%%%%%%%%%%%%%%%%%%%%%%%%%%%%%%%%%%%%%%%%%%%%
\section{Introduction}
\label{s:intro}

A stochastic background of gravitational radiation is usually defined
as a random gravitational-wave signal produced by a {\em large number
  of weak, independent, and unresolved sources}.  It can be of either
astrophysical or cosmological origin.  The signal is random in the
sense that it can be characterized only statistically, in terms of
expectation values of the Fourier components of a plane-wave expansion
of the metric perturbations.  For a sufficiently large number of
independent sources, the background will be Gaussian by the central
limit theorem.  Knowledge of the first two moments of the distribution
will then suffice to determine all higher-order moments, meaning that
the quadratic expectation values (or {\em covariance} matrix) of the
Fourier components completely define a Gaussian background of
gravitational radiation.  For non-Gaussian backgrounds, the only
difference is that the probability distribution of the Fourier
components is no longer Gaussian.  Thus, third and/or higher-order
moments of the distribution are now required.

Although there is general agreement with the above definition, there
has been some confusion and/or disagreement about some of the defining
properties of a stochastic background, in particular, related to the
resolvability of a signal and its relationship to duty-cycle
\cite{Rosado:2011, Regimbau-Mandic:2008, Regimbau-Hughes:2009, Regimbau:2011}.  
In order to avoid such
confusion in this paper, we give operational definitions for these
properties, framed in the context of Bayesian inference.  For
instance, we define a signal to be {\em stochastic} if it is more
parsimonious (in a Bayesian model selection sense) to search for that
signal using a stochastic signal model for the waveform than using a
deterministic signal model.  We also define a signal to be {\em
  resolvable} if it can be decomposed into {\em separate} (i.e.,
non-overlapping in either time or frequency) and {\em individually
  detectable} signals, again in a Bayesian model selection sense.
This definition of resolvability is more restrictive than that of
Rosado~\cite{Rosado:2011}, who defines a signal to be resolvable if it
is separable, independent of detectability.  With our definition, it
is possible to have separable signals that are not detectable (e.g.,
``subthreshold" low-duty cycle bursts \cite{Thrane:2013} or
non-overlapping low-SNR sinusoids), and signals that are detectable
but not resolvable (e.g., a Gaussian stochastic background integrated
over a large enough time or large enough frequency band).

In addition, for non-Gaussian backgrounds associated with the
superposition of signals from many astrophysical sources, there will
sometimes be cases where a few bright signals standout above the
lower-amplitude ``confusion" background.  These resolvable
deterministic signals should be `subtracted' from the data, leaving a
residual non-deterministic background whose statistical properties we
would like to determine.  In the context of Bayesian inference, this
`subtraction' is done by allowing {\em hybrid} signal models, which
consist of both parametrized deterministic signals and
non-deterministic backgrounds.  By using such models we can
investigate the statistical properties of the residual background
without the influence of the resolvable signals.  This is ultimately
the property of the stochastic background that we would like to
determine.

The closely related question of whether a population of astrophysical signals is more likely to
be first detected via a stochastic cross-correlation analysis or a template-based search for
individual signals has recently been considered by Rosado {\it et al.}~\cite{Rosado:2015epa} in the
context of pulsar timing detection of the low-frequency, slowly-evolving signals from binary
supermassive black holes, and by Mandel~\cite{mandel} in the context of ground-based interferometer
detections of chirping neutron star binaries and continuous waves from non-axisymmetric
spinning neutron stars. Rosado {\it et al.}~found that pulsar timing arrays are most likely to
first detect a stochastic background, though in some cases a bright and
nearby source may be detected first. Mandel concluded that individual signals will always be
detected first for non-overlapping, evolving signals, while a stochastic background may
be detected first for non-evolving, overlapping signals, and gave conditions for when this
might occur. Our results are broadly in agreement with these studies, though it is difficult to
directly compare our results since we frame the problem in different ways and have different
definitions for what it means for a signal to be considered deterministic or stochastic.

In this paper, we apply Bayesian inference to non-Gaussian
gravitational-wave backgrounds, which are produced whenever the
overlap of the gravitational-wave signals in time-frequency space is
sufficiently low that the central-limit theorem does not apply.
Previous analyses for non-Gaussian backgrounds have typically been
framed in the context of frequentist statistics, involving
modifications~\cite{Drasco-Flanagan:2003, Thrane:2013} of the standard
optimally-filtered cross-correlation
statistic~\cite{Allen-Romano:1999} used to search for Gaussian
backgrounds.  Here we use Bayesian inference to address the same
problem.  We apply Bayesian model selection to compare several
signal$+$noise models for simulated data consisting of uncorrelated
Gaussian detector noise plus a superposition of sinusoidal signals
from an astrophysical population of gravitational-wave sources.  The
analysis is done in the frequency domain since the signals we consider
are well localized in frequency and spread out in time, but our
results apply equally well to signals that are localized in time and
spread out in frequency, such as a population of burst signals
occurring with some Poisson rate.  For simplicity, we consider a pair
of co-located and co-aligned detectors with white detector noise, but
the method can be extended to more realistic detector configurations
and power spectra.

The general trend we observe from our simulations is that a
deterministic signal model is favored whenever the number of sources
contributing to the background is sufficiently small; a non-Gaussian
stochastic model is preferred for an intermediate number of sources;
and a Gaussian stochastic model is preferred for large source numbers.
However, due to large variations between individual signal
realizations, the boundaries between the three regimes are not sharply
defined.  We find that a hybrid, trans-dimensional model comprised of
a deterministic signal model for individual bright sources and a
Gaussian-stochastic signal model for the remaining confusion
background outperforms all other models in most instances.

The remainder of the paper is organized as follows: In
Sec.~\ref{s:bayesian-inference} we give a brief overview of Bayesian
inference, and apply it to the specific case of non-Gaussian
gravational-wave backgrounds in Sec.~\ref{s:method}.  There we define
the relevant noise and signal probability distributions, likelihood
functions, prior and posterior probability distributions, etc.~needed
for our analysis.  In Sec.~\ref{s:models} we define the various
signal$+$noise models that we use for the Bayesian model selection
calculations, the results of which, for simulated data, are described
in detail in Sec.~\ref{s:simulations}.  Finally, in
Sec.~\ref{s:discussion}, we discuss the relevance of the results in
the context of current searches for gravitational-wave backgrounds.
Appendix~\ref{a:bayesfactorapproxs} includes a discussion of different
approaches for calculating Bayes factors.

%%%%%%%%%%%%%%%%%%%%%%%%%%%%%%%%%%%%%%%%%%%%%%%%%
\section{Bayesian inference -- overview}
\label{s:bayesian-inference}

Bayesian inference is a powerful tool for assessing the plausibility
of hypotheses, given a set of observations and prior
information~\cite{Jaynes:2003}.
%Since observations and prior information are
%always incomplete and imprecise, the assessments 
%are necessarily probabilistic in nature.
It allows you to update your degree of belief in a particular
hypothesis, based on how well the hypothesis (or model) fits the
observed data.
It also implements a quantitative version of Occam's
razor~\cite{Jaynes:2003}, which says that given two models that fit
the data equally well, the simpler model should be preferred.  This
result falls naturally out of a Bayesian model selection calculation,
where one calculates the posterior odds ratio of one model against
another.  If two models fit the data equally well but have different
parameter space volumes, then the model with the larger parameter
space volume is penalized by the ratio of the larger parameter space
volume to the smaller volume.

Using Bayesian inference to analyze a particular problem is very
simple in principle---one applies Bayes' theorem, cf.~(\ref{e:bayes}),
to calculate posterior probability distributions given a likelihood
function (which specifies the probabilty of the data given the model
and the value of any parameters associated with it) and a prior
probability distribution for the model and its parameters.  In
practice, however, these calculations can be extremely
computationally-intensive, especially for models having a large number
of parameters.  But in recent years, thanks to advances in high-speed
computing and the development of efficient sampling
algorithms~\cite{Skilling:2006, importanceMultiNest2013}, integrations
over model parameter spaces having hundreds or even thousands of
dimensions are now possible.  Thus, the use of Bayesian inference to
solve diverse problems in the physical sciences has increased
dramatically, given the ability to do numerical calculations, which,
in the past, were not possible in practice.

In particular, in the field of gravitational-wave data analysis, it is
now common to see Bayesian inference used for: (i) detector noise
estimation and modeling~\cite{bayesline, bayeswave}, (ii)
sky-localization of signals from unmodeled graviational-wave bursts,
and (iii) parameter estimation for gravitational-wave signals
associated with many different sources---binary inspiral events,
continuous-wave sources (e.g., non-axisymmetric rotating neutron
stars), and stochastic gravitational-wave
backgrounds~\cite{vanHaasteren-et-al:2009} of either astrophysical or
cosmological origin.  Although Bayesian inference and frequentist
optimally-filtered statistic methods give equivalent results for
sufficiently simple signal and noise models and simple choices for the
priors, the Bayesian formalism allows one to more easily handle
problems involving more complicated models and/or non-trivial priors.
This is the case when the model contains so-called {\em nuisance
  parameters} (such as non-negligible correlated noise), which are not
of direct astrophysical interest, but nonetheless affect statistical
statements about the signal parameters.  For example, in the presence
of correlated noise, the standard optimally-filtered cross-correlation
statistic~\cite{Allen-Romano:1999} for isotropic gravitational-wave
backgrounds no longer corresponds to the optimal combination of the
data from the two detectors.  Calculating the maximum of the
likelihood function is more complicated for this case, with no
analytic closed-form solution in general.  But a Bayesian approach to
this problem, which numerically explores the likelihood function using
e.g., Markov Chain Monte Carlo (MCMC) methods, is a viable
alternative.

For gravitational-wave backgrounds generated by a superposition of
signals from a population of astrophysical sources, Bayesian inference
is particularly convenient since it allows one to compare several
viable signal$+$noise models.  Depending on the number of sources
emitting gravitational waves in a particular time-frequency volume,
the measured signal could be either: (i) stochastic and Gaussian
distributed, (ii) stochastic but non-Gaussian, (iii) a superposition
of individually resolvable signals, or (iv) some combination of both
deterministic resolvable signals and a non-deterministic (i.e.,
stochastic) background.  Using Bayesian model selection, we can rank
these various models, and thus characterize the gravitational-wave
component of the data.  The following sections describe this procedure
for the case of simulated data consisting of Gaussian white detector
noise plus a superposition of sinusoidal signals from an astrophysical
population of gravitational-wave sources.

%%%%%%%%%%%%%%%%%%%%%%%%%%%%%%%%%%%%%%%%%%%%
\section{Bayesian inference applied to non-Gaussian
backgrounds}
\label{s:method}

In this section we specify the various probability distributions,
likehood functions, posterior distributions, etc.~that we will need in
order to apply Bayesian inference to searches for non-Gaussian
gravitational-wave backgrounds.  Readers interested in more details
regarding some of the calculations performed in this section should
consult e.g., \cite{Cornish-Romano:2013}.

\subsection{Noise and signal probability distributions}

For simplicity, consider the simple case of $N$ samples
of data in a pair of co-located and co-aligned detectors:
\begin{equation}
{\mb s}_1 = {\mb n}_1 + {\mb h}\,,\quad 
{\mb s}_2 = {\mb n}_2 + {\mb h}\,,
\end{equation}
where ${\mb s}_1 =[s_{11}, s_{12}, \cdots  s_{1N}]^T$, etc.
We will assume that the noise in each detector is Gaussian, 
white, and independent of one another, with zero mean and 
variance $\sigma_1^2$, $\sigma_2^2$:
\begin{equation}
p_n({\mb n}\vert \vec\theta_n)  
= \frac{1}{\sqrt{{\rm det}(2\pi {\mb C}_n)} }\, 
e^{-\frac{1}{2}\, {\mb n}^T {\mb C}_n^{-1} {\mb n}}\,,
\label{noise}
\end{equation}
where
\begin{equation}
{\mb n} = \left[
\begin{array}{c}
{\mb n}_1\\
{\mb n}_2
\end{array}
\right]\,,
\quad
{\mb C}_n
=\left[
\begin{array}{cc}
\sigma_1^2\, \unit_{N\times N} & \zero_{N\times N} \\
\zero_{N\times N} & \sigma_2^2\,\unit_{N\times N}
\end{array}
\right]\,,
\end{equation}
and $\vec \theta_n =\{\sigma_1, \sigma_2\}$.  The signal ${\mb h}$,
which is common to both detectors, is assumed to come from a
probability distribution $p_h({\mb h}|\vec \theta_h)$, which need not
be Gaussian.  The probability distribution $p_h({\mb h}|\vec
\theta_h)$ is called a {\em parameterized signal prior} and
$\vec\theta_h$ are called {\em hyperparameters} \cite{Good:1965,
  Morris-Normand:1992}.  Examples of parameterized signal priors
include: 
\\ \\ 
(i) {\em Gaussian, white signal prior}:
\begin{equation}
\label{Gaussian_prior}
p_h({\mb h}\vert \vec\theta_h)
=\prod_{i=1}^N
\frac{1}{\sqrt{2\pi\sigma_h^2}} e^{-h_i^2/2\sigma_h^2}\,,
\end{equation}
where $\vec\theta_h=\{\sigma_h\}$.
\\
\\
(ii) {\em Two-component Gaussian, white signal prior}:
\begin{multline}
p_h({\mb h}\vert \vec\theta_h)=
\prod_{i=1}^N\left[
\xi\,\frac{1}{\sqrt{2\pi\alpha^2}}\, e^{-h_i^2/2\alpha^2}
\right.
\\
\left.
+(1-\xi)\,\frac{1}{\sqrt{2\pi\beta^2}}\, e^{-h_i^2/2\beta^2}
\right]\,,
\label{mixture_gaussian_prior}
\end{multline}
where $\vec\theta_h=\{\xi, \alpha, \beta\}$.
The two-component Gaussian signal prior 
reduces to the Gaussian 
signal prior in the limit $\xi\rightarrow 1$.
It reduces to the Drasco and Flanagan signal prior 
\cite{Drasco-Flanagan:2003} in the limit $\beta\rightarrow 0$.
The Drasco and Flanagan signal prior corresponds to 
Gaussian bursts with root-mean-square (rms) amplitude
$\alpha$ and probability $0\le \xi\le 1$.
\\
\\
(iii) {\em Non-standardized Student's $t$-distribution signal prior}:
\begin{equation}
\label{student_prior}
p_h({\mb h}\vert \vec\theta_h)=
\prod_{i=1}^N\left[
\frac{\Gamma\left(\frac{\nu+1}{2}\right)}{\alpha\,\Gamma\left(\frac{\nu}{2}\right)\sqrt{\pi\nu}}
\left(1+\frac{1}{\nu}\frac{h_i^2}{\alpha^2}\right)^{-\frac{\nu+1}{2}}
\right]\,,
\end{equation}
where
\begin{equation}
\Gamma(\nu) = \int_0^\infty dx\>x^{\nu-1}e^{-x}
\end{equation}
is the Gamma function and $\vec\theta_h=\{\nu, \alpha\}$.
The above distribution is an extension of the standard Student's 
$t$-distribution, which includes a scaling parameter $\alpha$ in 
addition to the number of degrees of freedom, $\nu>0$ (real).
(The $t$ of Student's $t$-distribution is given by $h_i/\alpha$.)
The scaling parameter $\alpha$ is related to the variance of each 
$h_i$  by
\begin{equation}
\sigma_h^2 = \alpha^2\,\frac{\nu}{\nu-2}\,,
\quad{\rm for\ } \nu>2\,.
\end{equation}
For $\nu\rightarrow\infty$, the non-standardized Student's 
$t$-distribution becomes a Gaussian distribution with the 
above variance.
\\
\\
(iv) {\em Multi-sinusoid signal prior}:
\begin{align}
&p_h({\mb h}|\vec\theta_h) = 
\delta\left({\mb h} - {\mb h}(\vec\theta_h)\right)\,,
\\
&h_i(\vec\theta_h) = \sum_{I=1}^M 
A_I \cos(2\pi f_I t_i - \varphi_I)\,,
\end{align}
where $i=1,2,\cdots, N$ and 
$\vec \theta_h = \{A_I, f_I, \varphi_I\vert I=1,2,\cdots, M\}$.
Here $M$ can take on any value between 0 and $M_{\rm max}$,
where $M_{\rm max}$ is the maximum number of allowed sinusoids
(e.g., $M_{\rm max}=100$).
This is a {\em deterministic} signal model corresponding 
to the superposition of $M$ individually resolvable sinusoids.
\smallskip

Although it is possible to write down more complicated non-Gaussian
signal priors (since there are an infinite number of ways for a signal
to be non-Gaussian), for the analysis considered in this paper, we
will restrict ourselves to those given above.

%%%%%%%%%%%%%%%%%%%
\subsection{Likelihood functions}

To construct the likelihood function, we first adopt a waveform
template ${\mb h}$ and form the residuals 
${\mb r}_1={\mb s}_1-{\mb h}$ and ${\mb r}_2={\mb s}_2-{\mb h}$.  
We demand that the residuals
be consistent with the probability distribution for the noise
(cf.\ (\ref{noise})), which gives rise to a multivariate Gaussian
likelihood function for the data:
\begin{equation}\label{temp}
p({\mb s}\vert \vec\theta_n, {\mb h}) \equiv 
p_n({\mb r}\vert \vec\theta_n) =
\frac{1}{\sqrt{{\rm det}(2\pi {\mb C}_n)}}\, 
e^{-\frac{1}{2}\, {\mb r}^T {\mb C}_n^{-1} {\mb r}}\,,
\end{equation}
where 
\begin{equation}
\mb{s}=
\left[
\begin{array}{c}
{\mb s}_1\\
{\mb s}_2
\end{array}
\right]\,,
\quad
\mb{r}=
\left[
\begin{array}{c}
{\mb s}_1-{\mb h}\\
{\mb s}_2-{\mb h}
\end{array}
\right]\,.
\end{equation}
But since ${\mb h}$ are random variables for stochastic
signal models or specified functions of the parameters
$\vec\theta_h$ for deterministic signal models, 
we are not interested in the particular values of ${\mb h}$, 
but rather in the values of the parameters $\vec\theta_h$ 
that define the signal prior $p_h({\mb h}|\vec \theta_h)$.
We thus {\em marginalize} over ${\mb h}$ by performing the integral:
\begin{equation}
\label{marginalize}
p({\mb s}|\vec\theta)\equiv
p({\mb s}|\vec\theta_n,\vec\theta_h)=
\int d{\mb h}\>
p({\mb s}\vert \vec\theta_n, {\mb h})\, p_h({\mb h}\vert \vec\theta_h)\,.
\end{equation}
Here $\vec\theta \equiv \{\vec\theta_n, \vec\theta_h\}$ denotes the combined
set of noise and signal parameters.
\\
\\
(i) For the Gaussian signal prior, we find:
\begin{equation}
p({\mb s}|\vec\theta)
=\frac{1}{\sqrt{{\rm det}(2\pi {\mb C}_s)}}\, 
e^{-\frac{1}{2}\, {\mb s}^T {\mb C}_s^{-1}{\mb s}}\,,
\label{corr}
\end{equation}
where 
\begin{equation}
{\mb C}_s
={\mb C}_n + \sigma_h^2\,
\left[
\begin{array}{cc}
\unit_{N\times N} & \unit_{N\times N} \\
\unit_{N\times N} & \unit_{N\times N}
\end{array}
\right]\,.
\label{Cs}
\end{equation}
The likelihood function given by (\ref{corr}) and (\ref{Cs})
has the standard form used as the starting point for cross-correlation 
analyses for Gaussian stochastic 
backgrounds~\cite{vanHaasteren-et-al:2009, Adams-Cornish:2010}.
\\
\\
(ii) For the two-component Gaussian signal prior,
we obtain a two-component 
Gaussian distribution for the marginalized
likelihood, with covariance matrices similar to (\ref{Cs}),
but with $\sigma_h^2$ replaced by $\alpha^2$ 
and by $\beta^2$ for the two components, respectively:
\begin{multline}
p({\mb s}|\vec\theta)
=\xi\frac{1}{\sqrt{{\rm det}(2\pi {\mb C}_\alpha)}}\, 
e^{-\frac{1}{2}\, {\mb s}^T {\mb C}_\alpha^{-1}{\mb s}}
\\
+(1-\xi)\frac{1}{\sqrt{{\rm det}(2\pi {\mb C}_\beta)}}\, 
e^{-\frac{1}{2}\, {\mb s}^T {\mb C}_\beta^{-1}{\mb s}}\,,
\label{two-component}
\end{multline}
where 
\begin{equation}
{\mb C}_\alpha
={\mb C}_n + \alpha^2\,
\left[
\begin{array}{cc}
\unit_{N\times N} & \unit_{N\times N} \\
\unit_{N\times N} & \unit_{N\times N}
\end{array}
\right]
\end{equation}
and similarly for ${\mb C}_\beta$.
\\
\\
(iii) For the non-standardized Student's $t$-distribution, the 
marginalization integrals are all of the form:
\begin{equation}
\int_{-\infty}^\infty dh_i\>
e^{-\frac{1}{2}\frac{(s_{1i}-h_i)}{\sigma_1^2}}
e^{-\frac{1}{2}\frac{(s_{2i}-h_i)}{\sigma_2^2}}
\left(1+\frac{1}{\nu}\frac{h_i^2}{\alpha^2}\right)^{-\frac{\nu +1}{2}}\,.
\label{e:student-marg-int}
\end{equation} 
Unfortunately, we do not know how to analytically 
evaluate such an integral.
It is possible to consider an Edgeworth expansion of the 
Student $t$-distribution in terms of its 
non-zero cumulants, $c_2, c_4, \cdots$.
But then truncating the expansion after a finite number of terms 
would produce a {\em different} non-Gaussian distribution, that would 
behave differently in model comparision tests from the full 
Student's $t$-distribution.
Thus, if we want to use this distribution as one of our 
non-Gaussian signal models, we would need to evaluate 
the above integrals numerically.
\\
\\
(iv) For the deterministic multi-sinusoid signal model, the marginalized
likelihood is simply
\begin{equation}\label{temp}
p({\mb s}\vert \vec\theta) 
=
\frac{1}{\sqrt{{\rm det}(2\pi {\mb C}_n)}}\, 
e^{-\frac{1}{2}\, 
({\mb s}-{\mb h}(\vec\theta_h))^T {\mb C}_n^{-1} 
({\mb s}-{\mb h}(\vec\theta_h))}\,.
\end{equation}
%

%%%%%%%%%%%%%%%%%%%%%%%
\subsection{Posterior distributions and Bayesian model selection}
\label{s:modelselection}

Given the marginalized likelihood function 
$p({\mb s}\vert \vec \theta)$ and a prior probability 
distribution $\pi(\vec\theta)$ for the noise and 
signal parameters $\vec\theta\equiv\{\vec\theta_n,\vec\theta_h\}$, 
we can use Bayes' theorem
\begin{equation}
p(\vec\theta\vert {\mb s}) = 
\frac{p({\mb s}\vert \vec\theta)\pi(\vec\theta)}
{\int d\vec\theta'\>
p({\mb s}\vert \vec\theta')\pi(\vec\theta')}
\label{e:bayes}
\end{equation}
to calculate the joint posterior probability 
distribution for the parameters.
The posterior distribution for a 
{\em subset} of the parameters is obtained by 
marginalizing over the 
other signal and noise parameters.
For example, for the Gaussian signal model, 
the posterior distribution for the 
signal parameter $\sigma_h$ is obtained by evaluating
the following integral:
\begin{equation}
p(\sigma_h\vert {\mb s}) = 
\int d\sigma_1 \int d\sigma_2\>
p(\sigma_1, \sigma_2,\sigma_h\vert {\mb s})\,.
\end{equation}
Similar integrals will give the posterior 
distributions for $\sigma_1$ and $\sigma_2$.

In a similar manner, we can calculate
the posterior probability distribution 
for a signal$+$noise model ${\cal M}$
using Bayes' theorem in the form
\begin{equation}
p({\cal M}\vert {\mb s}) = 
\frac{p({\mb s}\vert {\cal M})\pi({\cal M})}
{\sum_I
p({\mb s}\vert {\cal M}_I)\pi({\cal M}_I)}\,.
\end{equation}
The quantity $p({\mb s}\vert{\cal M})$ is called 
the {\em evidence} for model ${\cal M}$.
It is just the likelihood function
$p({\mb s}\vert \vec\theta,{\cal M})$ 
marginalized over the parameter values
\begin{equation}
p({\mb s}\vert {\cal M})
= 
\int d\vec\theta\>
p({\mb s}\vert \vec\theta, {\cal M})
\pi(\vec\theta\vert {\cal M})\,,
\end{equation}
where we have explicitly indicated the model
dependence of both the prior and likelihood 
function.

To compare two models ${\cal M}_I$ and ${\cal M}_J$,
we simply take the ratio of the posterior 
probability distributions for these two models:
\begin{equation}
\frac
{p({\cal M}_I\vert {\mb s})}
{p({\cal M}_J\vert {\mb s})}
=
\frac
{p({\mb s}\vert {\cal M}_I) \pi({\cal M}_I)}
{p({\mb s}\vert {\cal M}_J) \pi({\cal M}_J)}\,.
\end{equation}
Note that the the common factor
$\sum_I p({\mb s}\vert {\cal M}_I)\pi({\cal M}_I)$
has canceled out when forming the ratio.
The left-hand side of the above equation
is the {\em posterior odds ratio} for model 
${\cal M}_I$ relative to ${\cal M}_J$;
we see from this equation that it equals the 
prior odds ratio times the ratio of the evidences.
This ratio of evidences is called the 
{\em Bayes factor} and is denoted by
\begin{equation}
{\cal B}_{IJ}({\mb s}) \equiv
\frac
{p({\mb s}\vert {\cal M}_I)} 
{p({\mb s}\vert {\cal M}_J)}\,.
\end{equation}
In many circumstances there is no a~priori reason to prefer one model
over another (i.e., the prior odds ratio is unity), so for these cases
the posterior odds ratio is just the Bayes factor.  If we fix some
model, e.g., ${\cal M}_0$, and calculate the Bayes factors of all the
other models relative to ${\cal M}_0$, the model with the largest
Bayes' factor is the preferred model given the data.

Table~\ref{t:bayesfactors} gives a list of possible Bayes
factor values and their interpretation in terms of 
the evidence in favor of one model relative to another.
\begin{table}
\begin{tabular}{|c|c|c|}
\hline
${\cal B}_{IJ}(\mb s)$ & $2\ln {\cal B}_{IJ}(\mb s)$ &
Evidence for model ${\cal M}_I$ relative to ${\cal M}_J$ \\
\hline
$<1$ & $<0$ & Negative (supports model ${\cal M}_J$) \\
1--3 & 0--2 & Not worth more than a bare mention \\ 
3--12 & 2--5 & Positive \\ 
12--150 & 5--10 & Stong \\ 
$>150$ & $>10$ & Very strong \\ 
\hline
\end{tabular}
\caption{Bayes factors and their
interpretation in terms of the evidence in favor of one model
relative to the other.}
\label{t:bayesfactors}
\end{table}

%%%%%%%%%%%%%%%%%%%
\subsection{Comparison to maximum-likelihood analyses}

It is interesting to compare the Bayesian model selection calculation
discussed above to a maximum-likelihood frequentist analysis, e.g.,
that presented in \cite{Drasco-Flanagan:2003}.  There they construct a
detection statistic by maximizing the likelihood ratio for a
signal$+$noise model ${\cal M}_1$ to the noise-only model ${\cal M}_0$:
\begin{equation}
\Lambda_{\rm ML}({\mb s})
\equiv \frac
{\max_{\vec\theta_n} \max_{\vec\theta_h} p({\mb s}\vert\vec\theta_n,\vec\theta_h,{\cal M}_1)}
{\max_{\vec\theta_n'} p({\mb s}\vert \vec\theta_n',{\cal M}_0)}\,.
\end{equation}
The Bayes factor calculation also involves a ratio of
two quantities, but instead of {\em maximizing} over the 
parameters, we {\em marginalize} over the parameters:
\begin{align}
{\cal B}_{10}({\mb s})
&\equiv
\frac
{p({\mb s}\vert {\cal M}_1)} 
{p({\mb s}\vert {\cal M}_0)}
\\
&=
\frac
{\int d\vec\theta_n\int d\vec\theta_h\>
p({\mb s}\vert \vec\theta_n,\vec\theta_h,{\cal M}_1)\pi(\vec\theta_n,\vec\theta_h\vert {\cal M}_1)}
{\int d\vec\theta_n'\> p({\mb s}\vert \vec\theta_n',{\cal M}_0)\pi(\vec\theta_n'\vert {\cal M}_0)}\,.
\end{align}
These two expressions can be related to one another by using 
the Laplace approximation to individually approximate 
the evidences $p({\mb s}|{\cal M}_1)$ and $p({\mb s}|{\cal M}_0)$.
As shown in App.~\ref{a:bayesfactorapproxs},
\begin{equation}
p({\mb s}|{\cal M})
\simeq p({\mb s}|\vec\theta_{\rm ML})
\frac{\Delta V_{\cal M}}{V_{\cal M}}\,,
\end{equation}
where $\Delta V_{\cal M}/V_{\cal M}$ is the fraction of the 
parameter space volume for model ${\cal M}$ needed to fit the data, 
and $\vec\theta_{\rm ML}$ are the values of the particular 
model parameters that maximize the likelihood.
Thus,
\begin{equation}
{\cal B}_{10}({\mb s})
\simeq \Lambda_{\rm ML}({\mb s})
%\frac{\Delta V_1}{\Delta V_0}\frac{V_0}{V_1}\,,
\frac{\Delta V_1/V_1}{\Delta V_0/V_0}\,,
\end{equation}
which shows that the Bayes factor is proportional to the 
frequentist maximum-likelihood ratio.
The proportionality constant 
is basically the Occam's factor mentioned in 
Sec.~\ref{s:bayesian-inference}, which penalizes a model
if its parameter space volume is larger than necessary
to fit the data. 

%%%%%%%%%%%%%%%%%%%%%%%%%%%%%%%%%%%%%
\subsection{Signal-to-noise ratios}
\label{s:SNR}

One of the parameters that we will use to describe the
simulations in Sec.~\ref{s:simulations} 
is the ratio of the power in the injected 
signals to that of the detector noise.
For a stochastic gravitational-wave background 
described by the 
one-sided strain power spectral density $S_h(f)$, 
the expected signal-to-noise ratio of the 
optimally-filtered cross-correlation statistic
in a pair of detectors $I$, $J$ is given by~\cite{Thrane-Romano:2013}
\begin{equation}
{\rm SNR}^2\big|_{\rm stoch}
= \sqrt{2T}
\left[
\int_0^\infty df\>\frac{\Gamma_{IJ}^2(f)S_h(f)}{P_{n_I}(f)P_{n_J}(f)}
\right]^{1/2}\,,
\end{equation}
where $\Gamma_{IJ}(f)$ is the overlap function between
detectors $I$ and $J$ (see, e.g., \cite{Christensen:1992, Flanagan:1993}).
For a pair of identical, co-located and co-aligned detectors
the above expression simplifies to
\begin{equation}
{\rm SNR}^2\big|_{\rm stoch}
= \sqrt{2T}
\left[
\int_0^\infty df\>\frac{P_h^2(f)}{P_n^2(f)}
\right]^{1/2}\,,
\label{e:SNR-stoch}
\end{equation}
where $P_h(f) \equiv \Gamma_{II}(f)S_h(f)$ is the 
gravitational-wave power in a single detector.
We are assuming here that the signal power is weak
relative to the noise, so that the total power in
detector $I$ is given by
$P_I(f)\equiv P_h(f) + P_{n_I}(f)\approx P_{n_I}(f)$.

For a deterministic signal described by the strain 
response $\tilde h(f)$, it is often more 
convenient to work with the {\em matched-filter}
signal-to-noise ratio, which has expected value
\begin{equation}
{\rm SNR}^2\big|_{\rm det}
= 4\int_0^\infty df\>\frac{|\tilde h(f)|^2}{P_n(f)}
= 2T\int_0^\infty df\>\frac{P_h(f)}{P_n(f)}\,.
\label{e:SNR-det}
\end{equation}
For this case $P_h(f) = \frac{2}{T}|\tilde h(f)|^2$
is the one-sided power spectral density for the signal.
Note that for deterministic signals, the squared 
signal-to-noise ratio scales with the number of frequency
bins $N_{\rm bins}$, 
while for stochastic signals it scales like $\sqrt{N_{\rm bins}}$.

%%%%%%%%%%%%%%%%%%%%%%%%%%%%%%%%%%%%%%%%%%%%%%%%%
\section{Signal$+$noise models and priors}
\label{s:models}

We consider the following five models for describing the
signal and noise:

\begin{description}

\item[${\cal M}_0$] - Noise-only model:

This is a noise-only model, which assumes uncorrelated, white Gaussian
noise in the two detectors.  There are only two parameters for this
model, $\vec\theta = \{\sigma_1, \sigma_2\}$.  For our simulations,
the prior on the noise variances are flat between 0 and 10.

\item[${\cal M}_1$] - Noise plus Gaussian stochastic
model:

White Gaussian detector noise plus a white Gaussian gravitational-wave
background.  There is one additional parameter corresponding to the
variance $\sigma_h^2$ of the background, so 
$\vec\theta = \{\sigma_1, \sigma_2, \sigma_h\}$.  
The prior on $\sigma_h^2$ is also between 0
and 10, just like the detector noise variances.

\item[${\cal M}_2$] - Noise plus non-Gaussian (two-component) 
stochastic model:

White Gaussian detector noise plus a white two-component Gaussian model 
for the gravitational-wave background.
There are three parameters for the two-component
Gaussian model: the variances 
$\alpha^2$ and $\beta^2$ for the two components, 
and the probability $\xi$ of one of the components.
(The probability of the other component 
necessarily equals $1-\xi$.)
Thus, $\vec\theta = \{\sigma_1, \sigma_2, \alpha, \beta, \xi\}$.
The prior on $\xi$ is flat from 0 to 1.
The prior on the variances are 0 to 10 for the 
wide component and 0 to 0.5 
on the narrow, delta-function-like component.

\item[${\cal M}_3$] - Noise plus deterministic multi-sinusoid signal model:

White Gaussian detector noise plus up to 100 deterministic sinusoids.
There are three parameters $\{A_I, f_I, \varphi_I\}$ corresponding to
the amplitude, frequency, and phase for each sinusoid.  Thus, for $M$
sinusoids, there are $2 + 3M$ pameters for this particular model 
$\vec\theta = \{\sigma_1,\sigma_2,A_I,f_I,\varphi_I\vert I=1,2,\cdots M\}$.
The prior on the amplitudes is uniform in the range $A\in[0,1000]$,
and the prior on the frequencies is uniform across range spanned by
the data.  The prior on the phases is uniform between 0 and $2\pi$.

\item[${\cal M}_4$] - Noise plus deterministic multi-sinusoid plus 
Gaussian background model:

White Gaussian detector noise plus a Gaussian gravitational-wave
background plus up to 100 sinusoids.
As for ${\cal M}_3$, there are three parameters (amplitude, 
frequency, and phase) for each sinusoid.
Thus, for $M$ sinusoids, there are $2 + 1 + 3M$ parameters 
for this model
$\vec\theta=\{\sigma_1,\sigma_2,\sigma_h,A_I,f_I,\varphi_I\vert I=1,2,\cdots M\}$.
The priors on the parameters are the same as in the
previous models.
This hybrid model allows us to effectively `subtract out' 
any sufficiently bright sinusoidal signals in the data.

\end{description}
Note that we do not consider a hybrid ``noise plus deterministic
multi-sinusoid plus {\em non-Gaussian} background" model in the 
above list, as we expect the subtraction of the bright sinusoids 
to remove most of the non-Gaussianity of the signal component.
Also, as we shall discuss further in Sec.~\ref{s:simulations},
we do not consider a signal$+$noise model with a 
non-standardized Student's $t$-distribution for the 
non-Gaussian stochastic gravitational-wave component.
This is because of the computational costs associated with the 
marginalized likelihood evaluations 
(see Eq.~(\ref{e:student-marg-int})),
which are needed for the Bayesian model selection calculations.

%%%%%%%%%%%%%%%%%%%%%%%%%%%%%%%%%%%%%%%%%%%%%%%%%
\section{Simulations}
\label{s:simulations}

\subsection{Astrophysical source populations}

Simulated data is generated by co-adding sinusoidal signals with
amplitudes drawn from one of three astrophysical models.  The
frequencies and phases of the sinusoids are drawn uniformly from the
prior ranges defined in the previous section.  Gaussian-distributed
noise with a white power spectrum is then added to the signals.  The
amplitude of the signals is scaled so as to produce a pre-specified
matched-filter signal-to-noise ratio (SNR) per frequency bin,
calculated as an average across all frequency bins, using
Eq.~(\ref{e:SNR-det}).

\begin{figure}[htbp]
\begin{center}
\includegraphics[width=.49\textwidth]{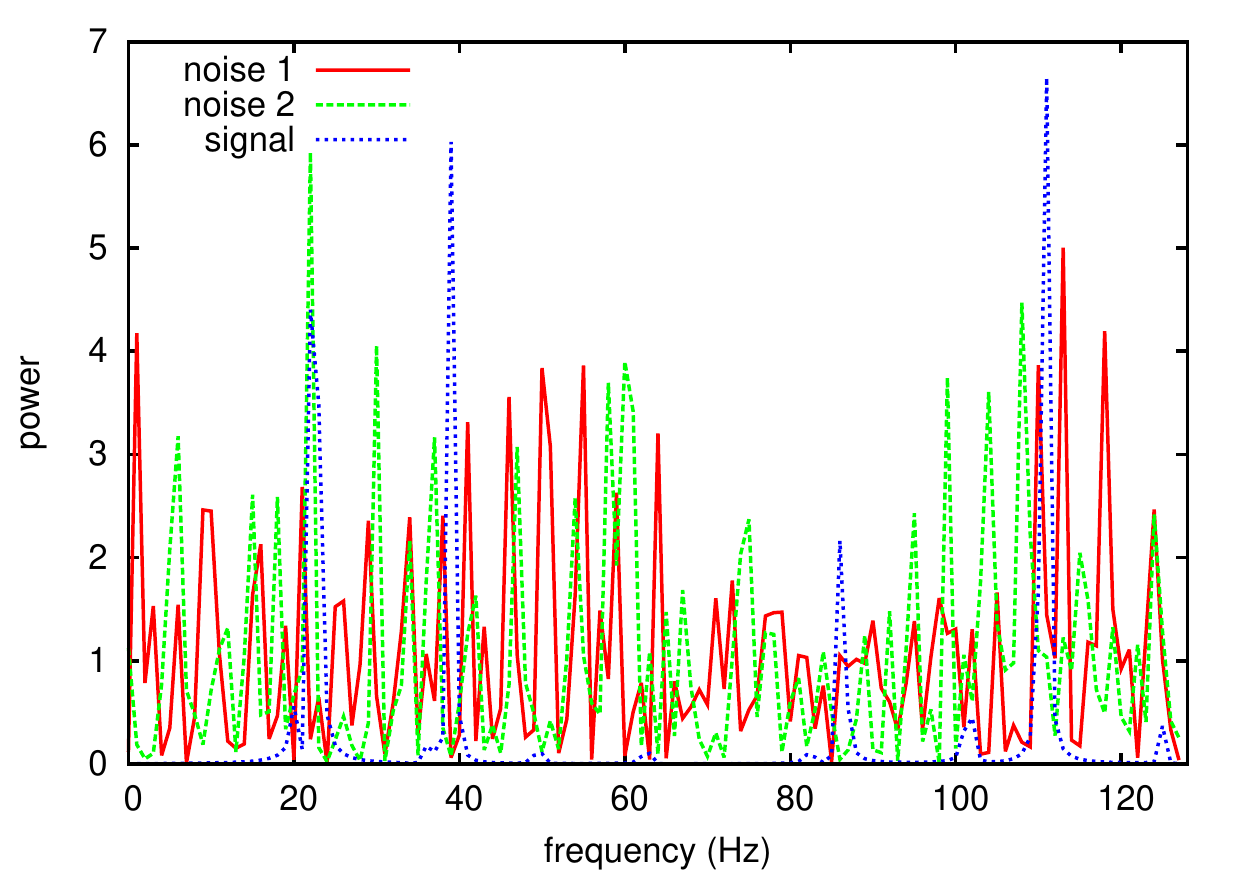}
\caption{The squared amplitude of the noise and signal components for
  data in two coincident and coaligned detectors consisting of white
  noise and a superposition of sinusoids drawn from an astrophysical
  source population (signal model 2), with a source density of 0.1/bin
  in 128 frequency bins and an average SNR-per-bin of 1. The SNR is
  dominated here by the four brightest sources.}
\label{f:signalnoise}
\end{center}
\end{figure}

Figure~\ref{f:signalnoise} is a plot of the squared amplitude of the
noise and signal components for a typical simulation using two
detectors.  For an SNR-per-bin of 1, the amplitudes of the
astrophysicals signals are of the same order-of-magnitude as the noise
in the two detectors, as can be seen in the figure.

We considered three astrophysical source models: Model 0 uniformly
distributes standard sirens (sources with the same intrinsic
amplitude) in space out to some cutoff radius $r=R$, after which the
density falls-off exponential with an e-folding scale of $0.25 R$.
Model 1 distributes standard sirens with a Gaussian distribution in
distance with density $\rho\propto e^{-r^2/2 R^2}$.  For model 0 the
number of sources in a spherical shell of radius $r$ is proportional
to $r^2$ out to $r=R$. For model 1 the number of sources in a
spherical shell of radius $r$ is proportional to the product $r^2
e^{-r^2/2R^2}$, and thus has a larger number of sources at smaller
$r$, as compared to the uniform distribution case.  Model 2 is based
on a population synthesis model for supermassive black hole binaries,
where the amplitude of the sources depends on both the mass of the
system and the distance. The usual frequency dependence of the
amplitude was artificially suppressed so as to produce a white
spectrum.

\begin{figure}[htbp]
\begin{center}
\includegraphics[width=.49\textwidth]{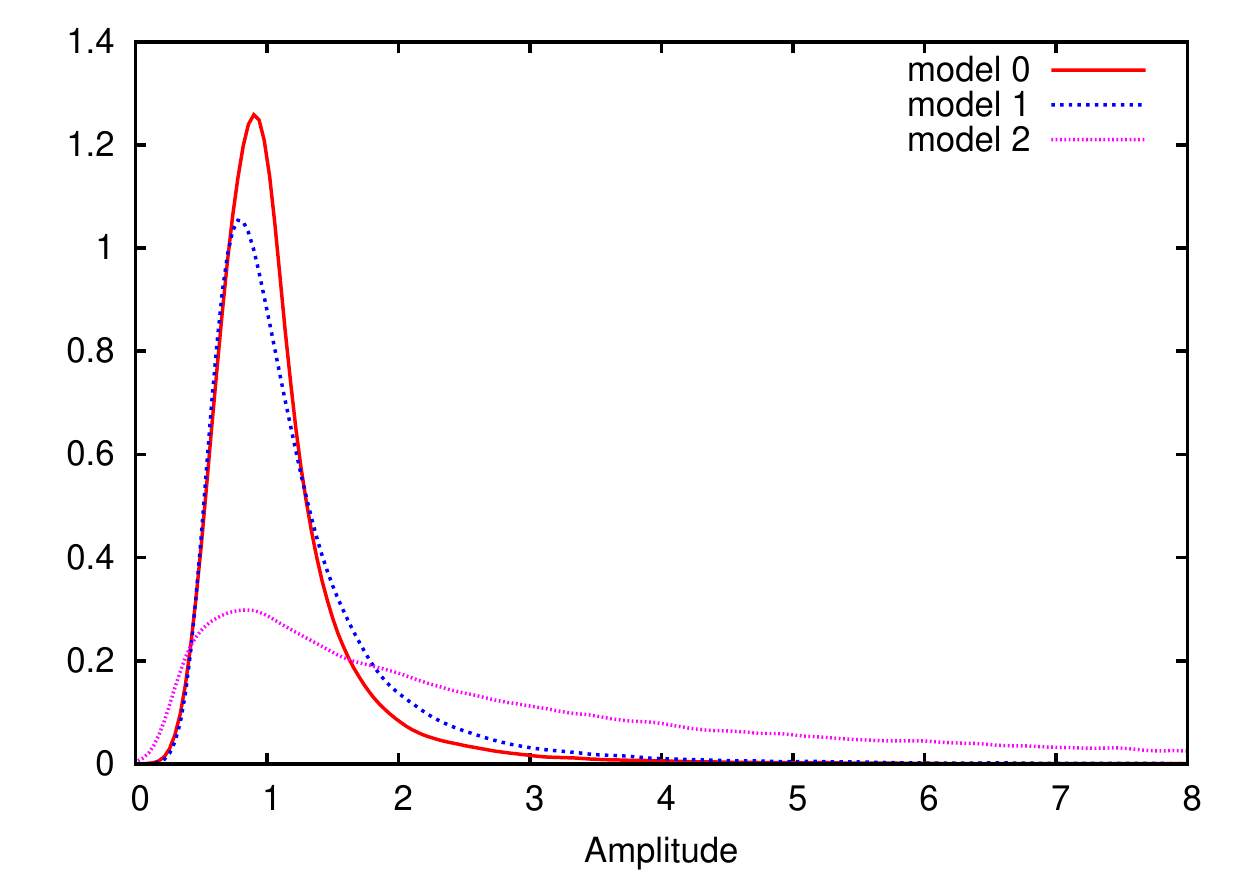}
\caption{Amplitude distributions for the three astrophysical source
  distributions considered in this study. The amplitude scale is
  arbitrary since the signal-to-noise ratios are set when producing
  simulated data sets drawn from these distributions. Models 0 and 1
  are for ``standard sirens'' (equal intrinsic amplitude sources) with
  different spatial distributions, while model 2 is based on a
  population synthesis model where some sources have much higher
  intrinsic amplitudes than others.}
\label{f:sourcemodels}
\end{center}
\end{figure}

The amplitude distributions for the three models are shown in
Fig.~\ref{f:sourcemodels}.  Models 0 and 1 have similar amplitude
distributions that are fairly tightly peaked, while model 2 has a
large tail extending to high amplitude.
%

%%%%%%%%%%%%%%%%%%%%%%%%%%%%%%%%%%%%%%%%%%
\subsection{Markov Chain Monte Carlo methods}
\label{s:MCMC}

We performed two types of analyses, both of which employed
trans-dimensional Reversible Jump Markov Chain (RJMCMC)
algorithms. The first type of analysis looked at the signal in a
single detector with no instrument noise. There the goal was to find
which of three statistical models best described the intrinsic
properties of the signal: a Gaussian distribution, a two-component
Gaussian distribution; or a non-standardized Student's
$t$-distribution.  A RJMCMC analysis extends the usual MCMC
exploration of the parameters of a single model to the exploration of
a range of models and their parameters, thus allowing us to produce
marginalized posteriors for both the model parameters {\em and}
posterior distributions for the relative probability that each model
is consistent with the data and our prior knowledge. In principle, a
single RJMCMC routine could explore all three probability
distributions at once, but we were able to achieve better mixing by
performing pair-wise comparisons between the Gaussian and
two-component Gaussian models and the Gaussian and the
non-standardized Student's $t$ model.  The ratio of the number of
iterations the Markov chain spends in each model yields Bayes Factors
between the Gaussian reference model and the two non-Gaussian
alternatives.

The second type of analysis considered the detection and
characterization of the astrophysical signals in the presence of
detector noise in a two-detector network. Here we considered the five
models described in Sec.~\ref{s:models}. Once again, a single RJMCMC
routine could simultaneously explore all five models, but achieving
efficient mixing between models with different parameterizations and
dimensionality is notoriously difficult. Instead, we again opted for a
pairwise approach, comparing the noise-only model ${\cal M}_0$ to each
of the four signal+noise models in turn. This yields a collection of
Bayes factors between the reference noise model and the four signal
models. It is important to note that models ${\cal M}_3$ and ${\cal
  M}_4$ are both complicated composite models that allow for a
variable number of sinusoids to be used in the model. Model ${\cal
  M}_4$ further allows for, but does not require, a Gaussian signal
component.  Thus ${\cal M}_4$ contains models ${\cal M}_3$, ${\cal
  M}_1$ and ${\cal M}_0$ as sub-cases.  If we had included a
two-component Gaussian in ${\cal M}_4$ then we would have been able to
explore all four signal models at once. We did check that the relative
probabilities for the sub-models included in ${\cal M}_4$ were
consistent with the relative probabilities found in the pair-wise
comparisons, though the larger model space did lead to larger
uncertainties on the Bayes factors.  The uncertainties were computed
from the variance of the running Bayes factors, and compared to
analytic estimates based on the number of transitions between the
models~\cite{bayeswave}.  Both methods yielded consistent error
estimates. We further checked that the error estimates were consistent
with the spread seen when repeating the analysis dozens of times with
different random number seeds.

%%%%%%%%%%%%%%%%%%%%%%%%%%%%%%%%%%%%%%%%%%
\subsection{Classifying the signals}

We begin looking at the statistical properties of the signals themselves. 
While this is not something we can do with actual observations
where the signals must be extracted from noisy data, 
it is interesting to compare the intrinsic properties of the signals
to the observed properties of the signals.
\begin{figure}[htbp]
\begin{center}
\includegraphics[width=.49\textwidth]{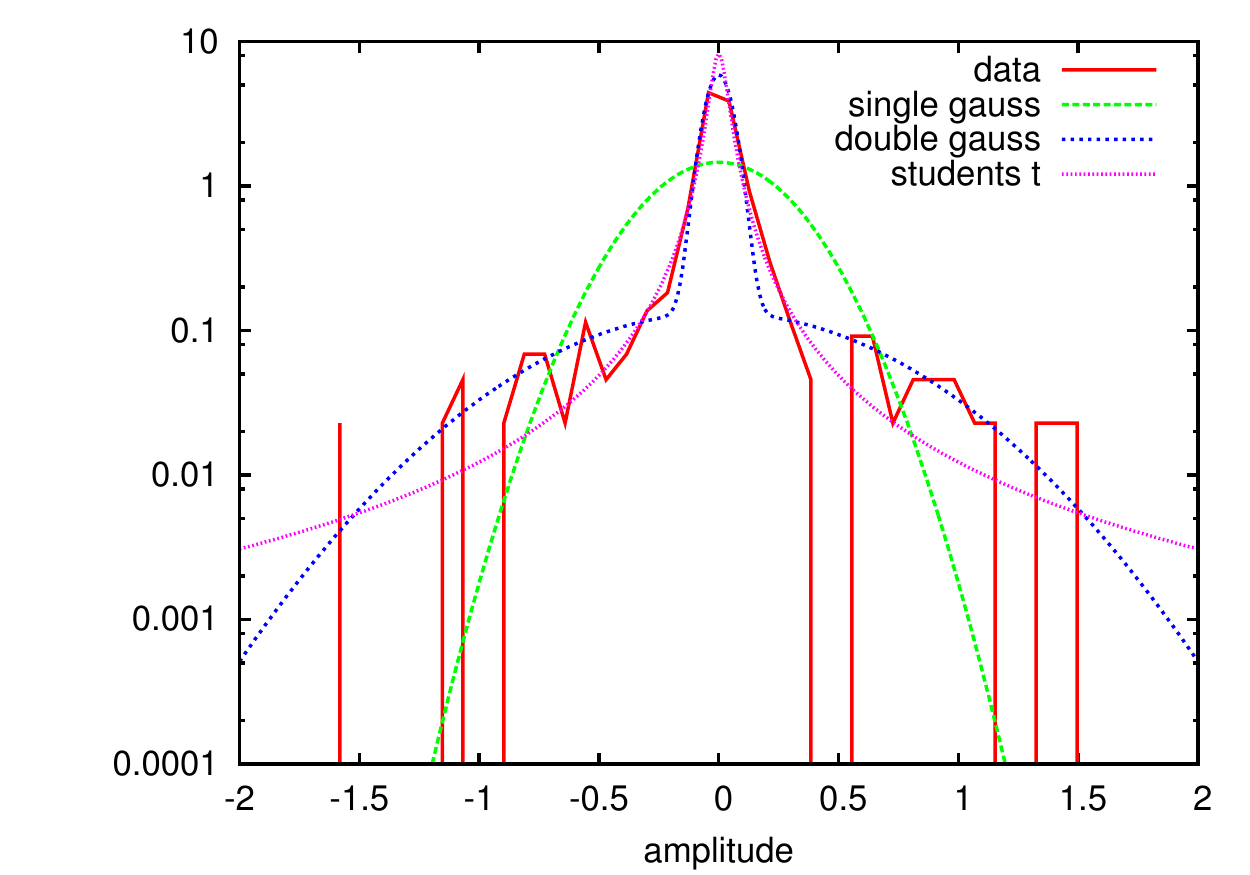}
\caption{Histogram of signal samples and the corresponding best fit
  single-Gaussian, double-Gaussian, and non-standardized Student's
  $t$-distribution for a signal consisting of a superposition of
  sinusoids drawn from an astrophysical population, with a source
  density of 0.1/bin in 256 frequency bins and an SNR-per-bin of 1.}
\label{f:histfit}
\end{center}
\end{figure}

Figure~\ref{f:histfit} is a histogram of signal samples as well as the
best fit Gaussian (``single gauss"), two-component Gaussian (``double
gauss"), and non-standardized Student's $t$-distribution for a
simulated signal with an average density of one source per ten
frequency bins.  As expected for such a sparse population, the
single-Gaussian fit is extremely poor compared to the two-component
Gaussian or non-standardized Student's $t$-distibution fit.

\begin{figure}[htbp]
{\includegraphics[width=.47\textwidth]{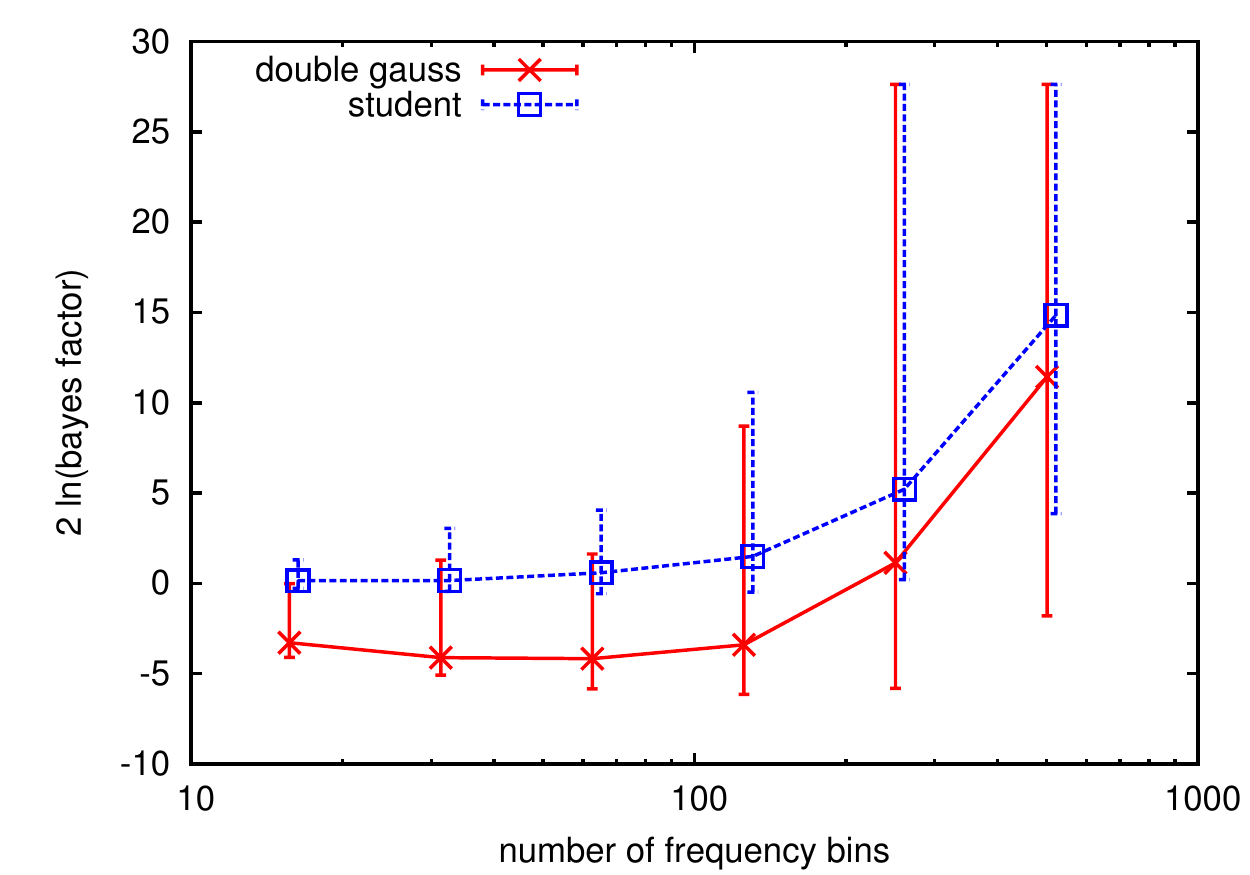}}
{\includegraphics[width=.47\textwidth]{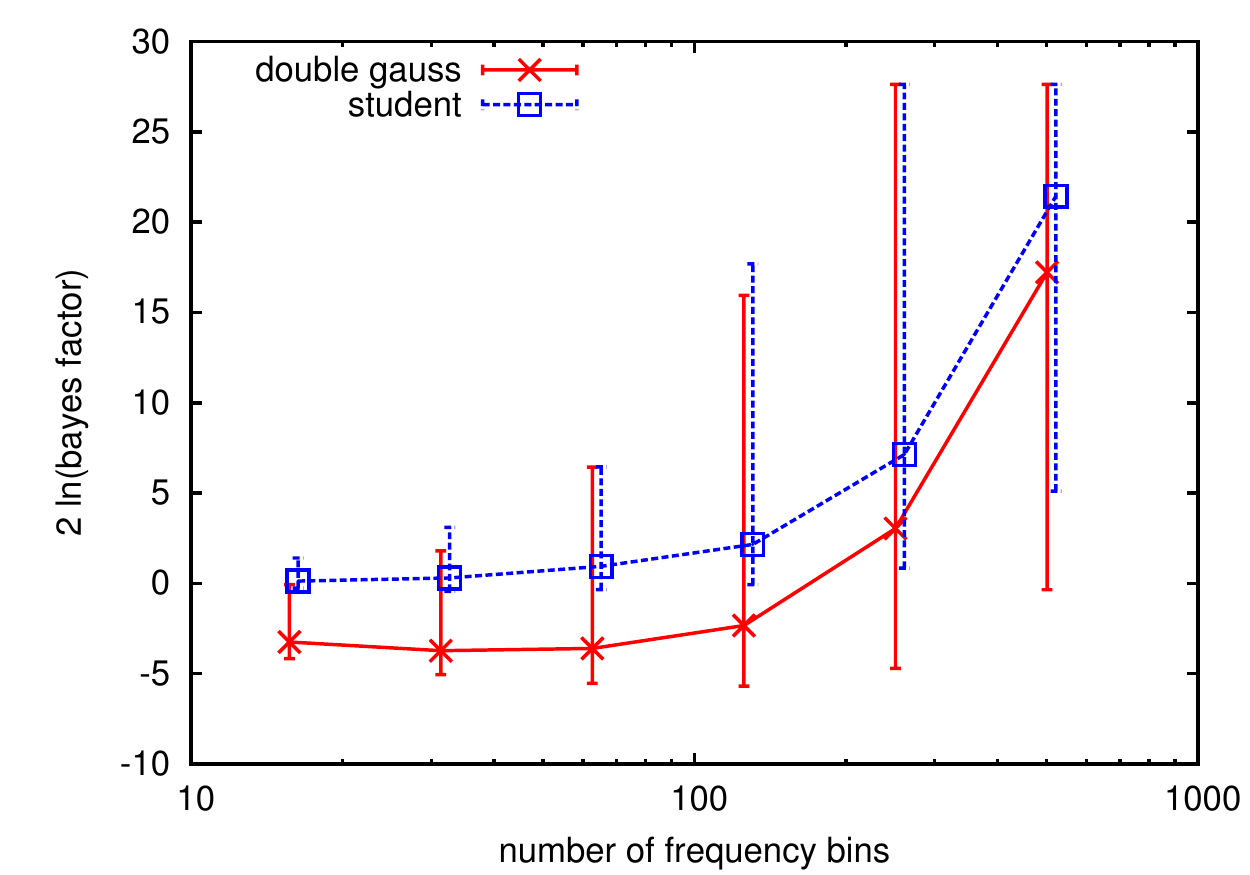}}
\caption{Bayes factor 80\% quantile intervals for the two-component
  Gaussian and non-standardized Student's $t$-distribution signal
  models as a function of the total number of frequency bins.  The
  source density was set to 10/bin for all the simulations.  In the
  upper panel the astrophysical sources are drawn from the source
  model 0. In the lower panel the astrophysical sources are drawn from
  source model 1.}
\label{f:errorbars_nbins}
\end{figure}
Figure~\ref{f:errorbars_nbins} is a plot of Bayes factor quantile
intervals as a function of the total number of frequency bins,
comparing the two-component Gaussian and non-standardized Student's
$t$-distribution models to the reference Gaussian model.  The source
density was set to 10/bin for these simulations.  The two panels
correspond to astrophysical source models 0 and 1.  In the upper panel
the astrophysical sources are drawn from source model 0.  In the lower
panel, the astrophysical sources are drawn from source model 1.  There
is no detector noise in these simulations.  Note that the Bayes
factors in the lower panel are shifted slightly higher relative to
those in the upper panel, consistent with the expectation that the
Gaussian-distributed astrophysical source population will tend to
produce closer---and hence more-easily resolvable---sources.  We see
that at this relative high source density, we need a large amount of
data (many frequency bins) to detect the subtle departure from
Gaussianity.

We should emphasize that the quantile intervals shown in
Fig.~\ref{f:errorbars_nbins} (and in several other figures to follow)
define the probability distribution for the Bayes factor values as
estimated from 256 independent realizations of the simulated signal
and noise for each set of parameter values: {\em these are not error
  bars on the individual Bayes factors}.  For a single realization of
the simulated signal and noise, the uncertainty in the value of the
Bayes factor as estimated from 128 independent Monte Carlo simulations
is $\lesssim 10\%$, which we can ignore in the quantile plots.

\begin{figure}[htbp]
\begin{center}
{\includegraphics[width=.49\textwidth]{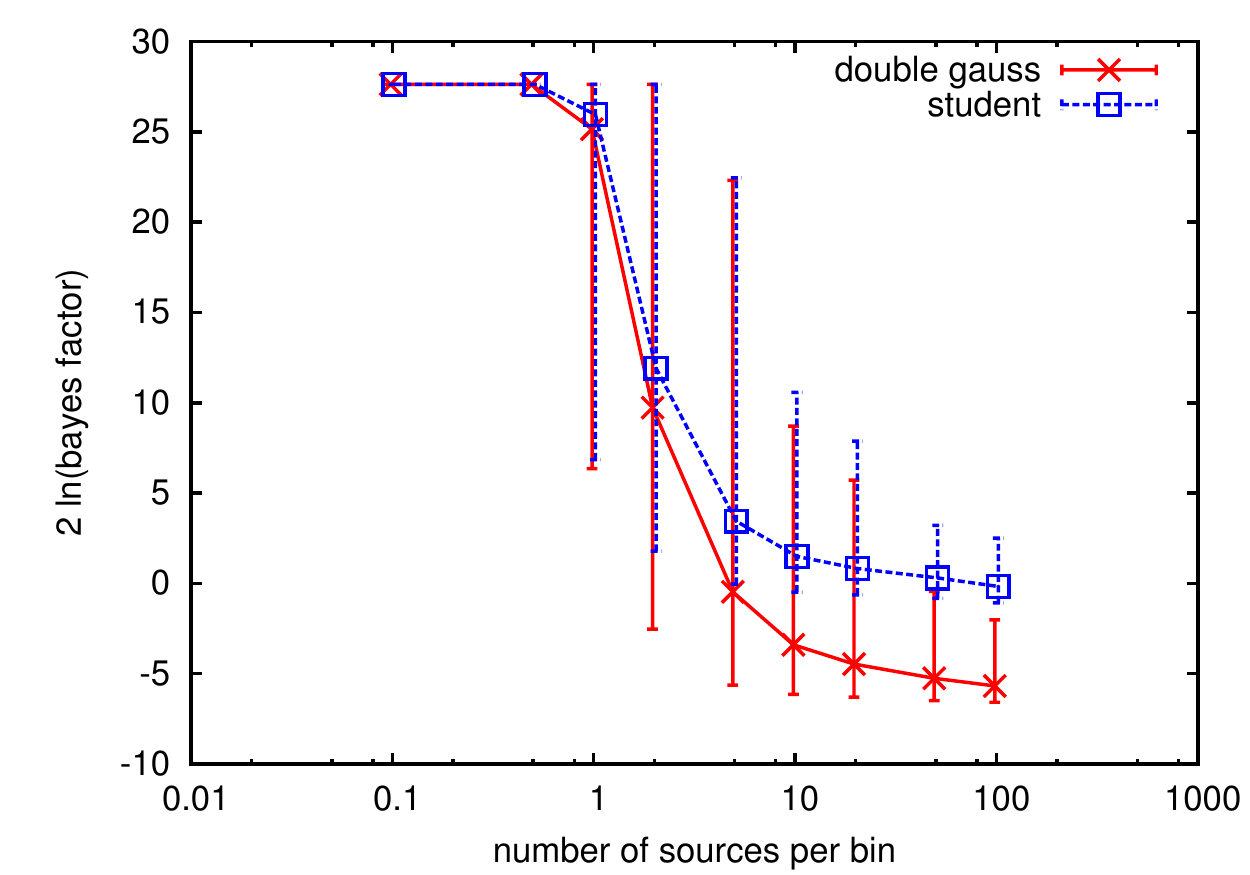}}
\caption{Bayes factor 80\% quantile intervals for the two-component
  Gaussian and non-standardized Student's $t$-distribution signal
  models as a function of the total number of sources per bin.  The
  total number of bins was set to 128 and the astrophysical sources
  were drawn from source model 0.}
\label{f:errorbars_Sden}
\end{center}
\end{figure}

Figure~\ref{f:errorbars_Sden} is a similar plot of Bayes factor
quantile intervals as a function of the number of sources per bin,
comparing the two-component Gaussian and non-standardized Student's
$t$-distribution.  The total number of bins was set to 128 and the
signals were drawn from astrophysical source model 0.  As expected
from the central limit theorem, the simulated data is consistent with
a Gaussian probability distribution when the average number of signals
per frequency bin is large. It is interesting to note the large spread
in the Bayes factors in the transition region between 1 and 10 sources
per bin. This tells us that some realizations look Gaussian, while
others look highly non-Gaussian.

Similar to the results shown in Figure~\ref{f:errorbars_nbins}, the
non-standardized Student's $t$-distribution model has consistently
higher Bayes factors than the two-component Gaussian model.  This
suggests using it over the two-component Gaussian model when modeling
non-Gaussian stochastic signals.  However, the fact that we are not
able to find an analytic expression for the corresponding marginalized
likelihood function (see Eq.~(\ref{e:student-marg-int})) means that
the Student's $t$-distribution model has a much higher computational
cost than the two-component Gaussian model.  As such, for all
subsequent model comparison simulations that we do---which include
simulated noise in a two-detector network---we use the two-component
Gaussian stochastic model instead of the more expensive
non-standardized Student's $t$-distribution model.

\subsection{Detecting and characterizing signals in noisy detector data}

Next we turn our attention to the observed properties of the signals
in a more realistic setup that includes instrument noise and a network
with two co-aligned and co-located detectors.  A multi-detector
analysis is needed to distinguish signals from noise.  In this study
we need to consider the dependence on SNR, in addition to the
dependence on source density and data volume (number of frequency
bins). In the infinite SNR limit we recover the pure signal analysis
described in the previous section.  For more realistic SNRs, a signal
that is best described as deterministic or non-Gaussian in the absence
of noise may favor a stochastic or Gaussian description in the
presence of noise when model simplicity wins out over model fidelity.

\begin{figure}[htbp]
\begin{center}
\includegraphics[width=.49\textwidth]{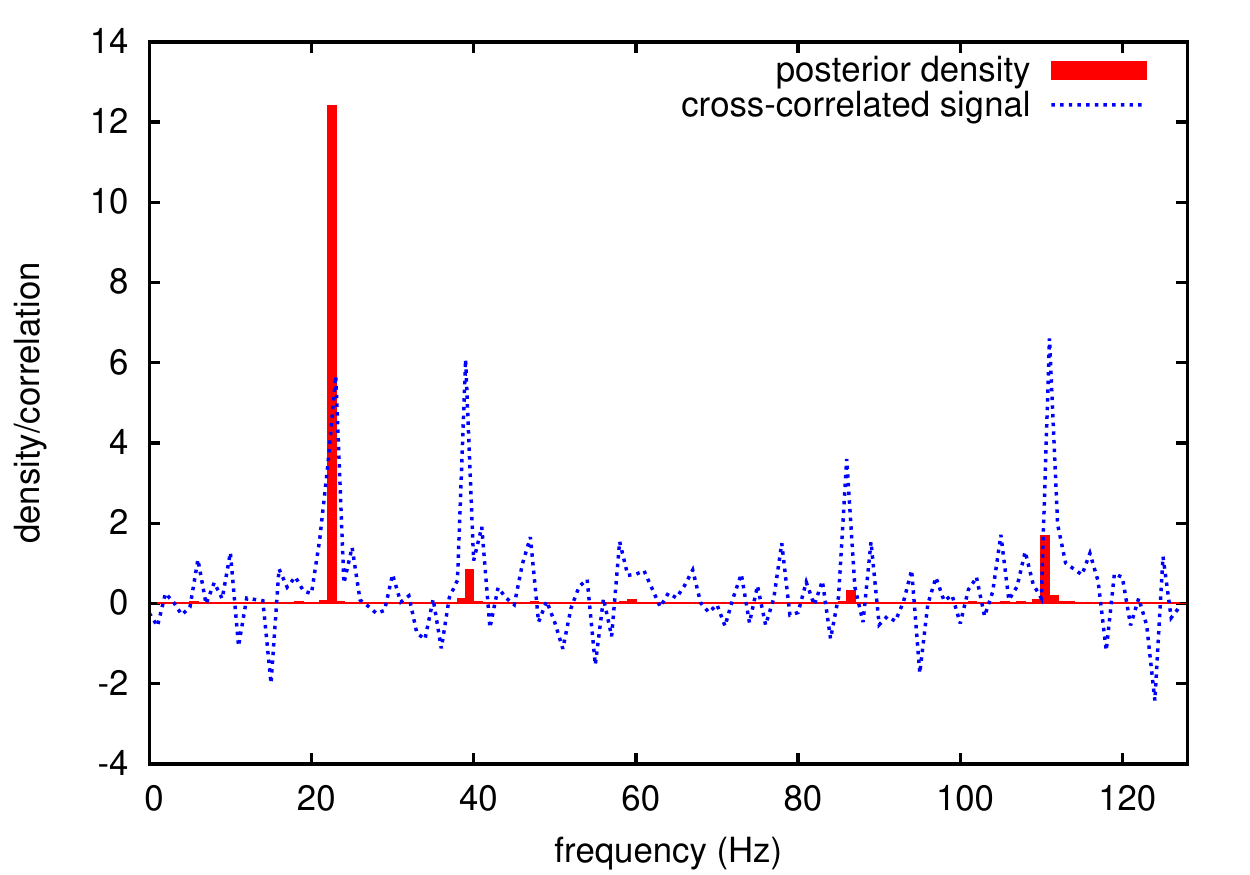}
\caption{The cross-correlated signal+noise in the two detectors for
  the simulated data shown in Figure 1 is compared to the scaled
  posterior density for the frequencies of the sinusoids found by a
  trans-dimensional MCMC analysis of the data. The three brightest
  signals in the data had amplitudes and frequencies 
$(A=4.46,\, f =  22.45)$, $(A = 4.41,\, f = 110.66)$ and 
$(A = 3.73,\, f =  39.23)$. 
  Only the brightest of these was a clear detection, though
  the analysis did occasionally lock onto the other signals.}
\label{f:sinusoids_search}
\end{center}
\end{figure}

In addition to the Gaussian and two-component Gaussian signal models
(${\cal M}_1$ and ${\cal M}_2$), we additionally consider a
deterministic model made up of the sum of sinusoids (${\cal M}_3$),
and a hybrid model with a Gaussian-stochastic component and a
collection of sinusoids (${\cal M}_4$).
Figure~\ref{f:sinusoids_search} compares the cross-correlated data in
two detectors to a marginalized posterior distribution for the
frequencies used by the multi-component sinusoid model. In this
instance, the brightest sinusoid in the data was confidently detected,
as indicated by the large peak in the posterior distribution. The
second and third brightest signals in the data were marginally
detected, as indicated by the secondary peaks in the posterior
distribution.

The question of wether the data is best described by a deterministic
model, a non-Gaussian stochastic model or a Gaussian-stochastic model
depends on many factors, including the source density, the noise
level, the number of frequency bins and the SNR-per-bin. In what
follows we explore the impact of each of these factors.

\begin{figure}[h]
{\includegraphics[width=.49\textwidth]{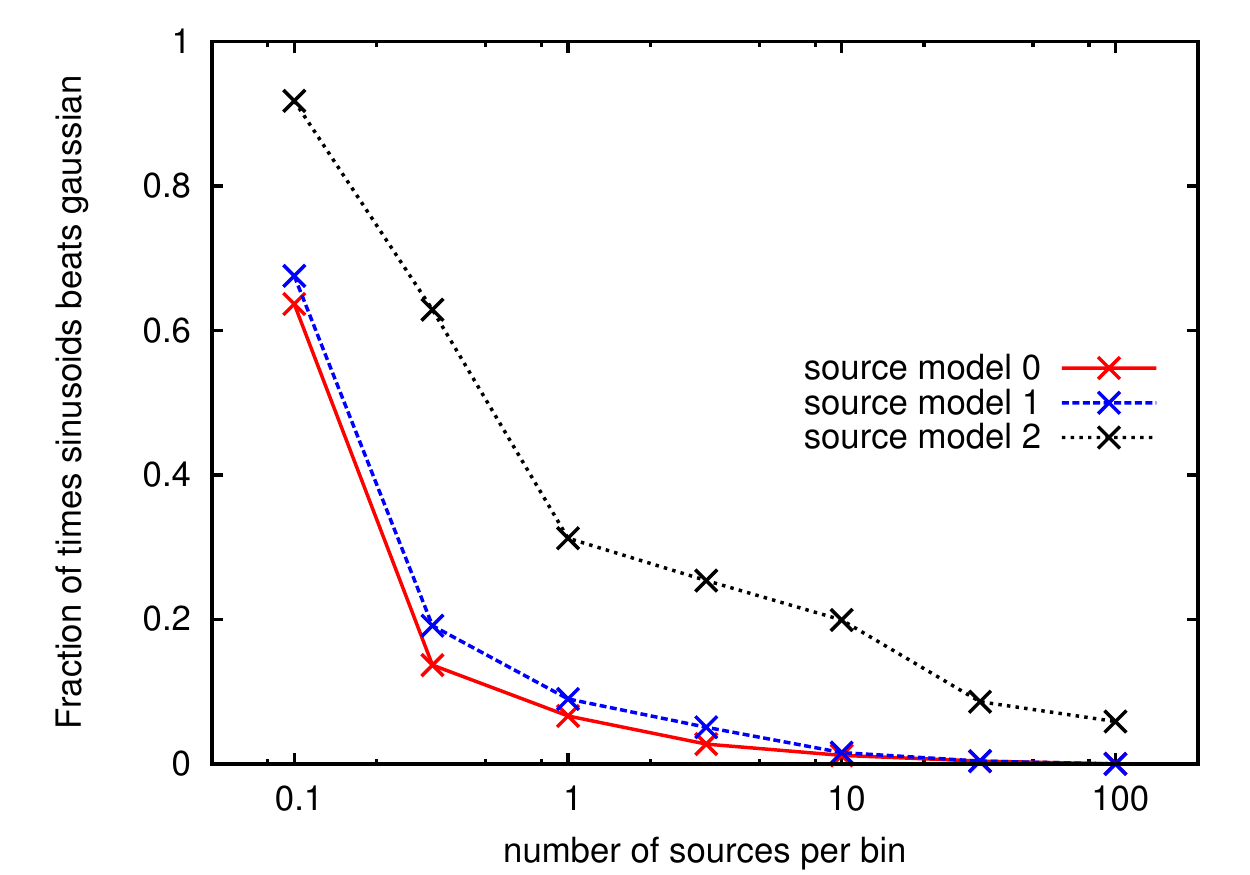}}
{\includegraphics[width=.49\textwidth]{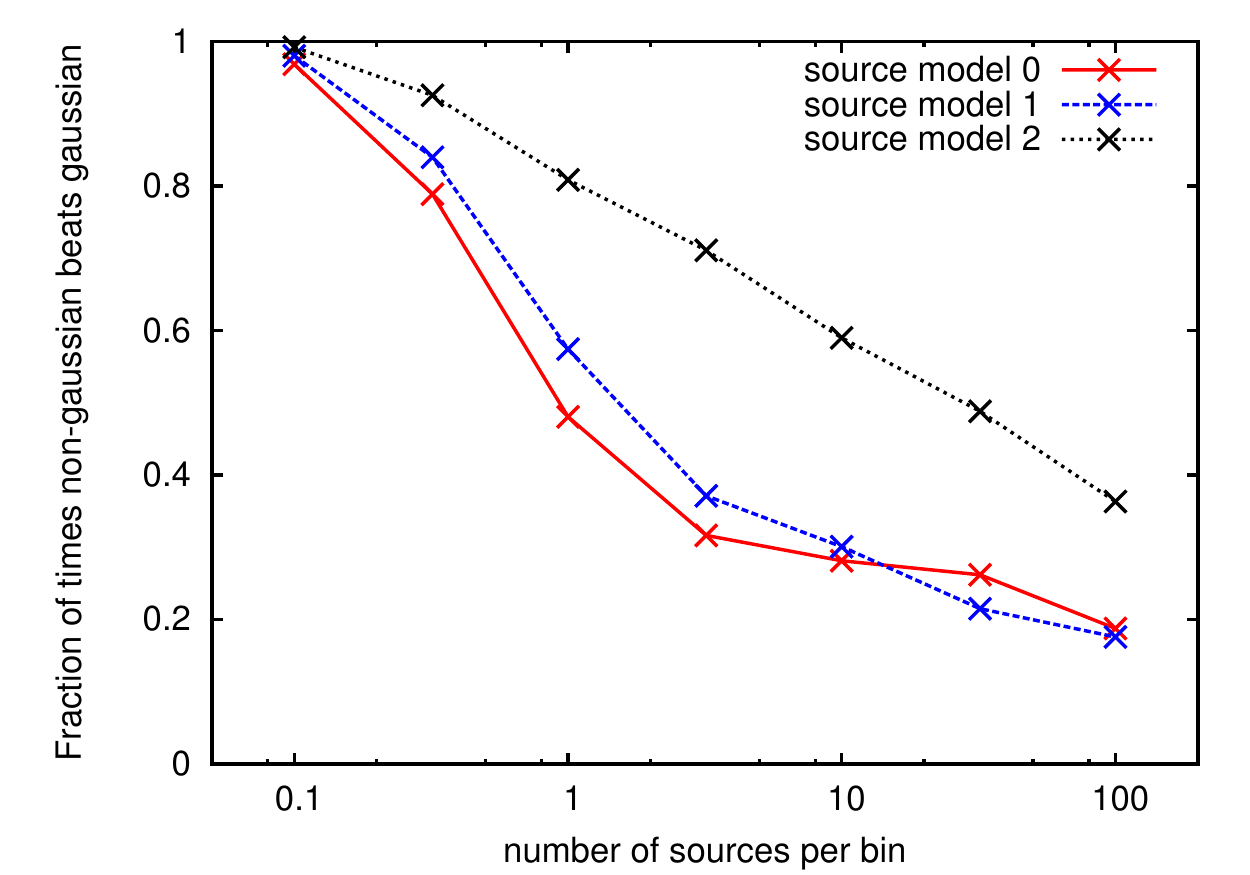}}
\caption{The fraction of times a non-Gaussian model has higher
  evidence than the Gaussian model as a function of the source density
  for the three source models.  For these simulations the number of
  bins was fixed at 32 and the average SNR-per-bin was fixed at 2.
  Upper panel: The fraction of times the deterministic, multi-sinusoid
  model ${\cal M}_3$ had higher evidence than the Gaussian-stochastic
  model ${\cal M}_1$.  Lower panel: The fraction of times the
  stochastic two-component Gaussian model ${\cal M}_2$ had higher
  evidence than the Gaussian-stochastic model ${\cal M}_1$.}
\label{f:SdenNoiseSourceModelAll}
\end{figure}

A major challenge that we face is that the outcomes vary greatly from
one simulation to the next, especially in the limit that there are few
sources and/or few frequency bins.  To counter this we performed a
large number of simulations and aggregated the results. When showing
Bayes factors between the various signal models and the noise model,
we display the mean values along with the 80\% quantile intervals
derived from the ensemble of simulations. More directly, we also
report the fraction of times each model had the highest Bayesian
evidence on a realization-by-realization basis. While the general
trend is that the Gaussian model is more likely to be favored as the
number of sources per bin increases, the deterministic model can
sometimes be preferred at high source density, and the
Gaussian-stochastic model can sometimes be preferred at low source
density. Note that while ground and space based interferometric detectors
and pulsar timing arrays nominally cover much larger frequency bands than we
consider here, their ``V'' shaped sensitivity curves limit the effective number of
frequency bins to 100's for interferometers and 10's for pulsar timing arrays.

Figure~\ref{f:SdenNoiseSourceModelAll} shows the fraction of times
that a non-Gaussian model has higher evidence than the Gaussian model
as a function of the source density for the three source models.  Here
the the number of bins was fixed at 32 and the average SNR-per-bin was
fixed at 2.  The general trend is as expected from the central limit
theorem---as the number of signals per frequency bin grows the data
looks less deterministic and more Gaussian. The more realistic source
model 2, which has a variety of intrinsic source luminosities, was
consistently less Gaussian than source models 0 and 1 which assumed
equal luminosity sources. Even at high source densities model 2 could
appear deterministic or non-Gaussian.

\begin{figure}[htbp]
{\includegraphics[width=.49\textwidth]{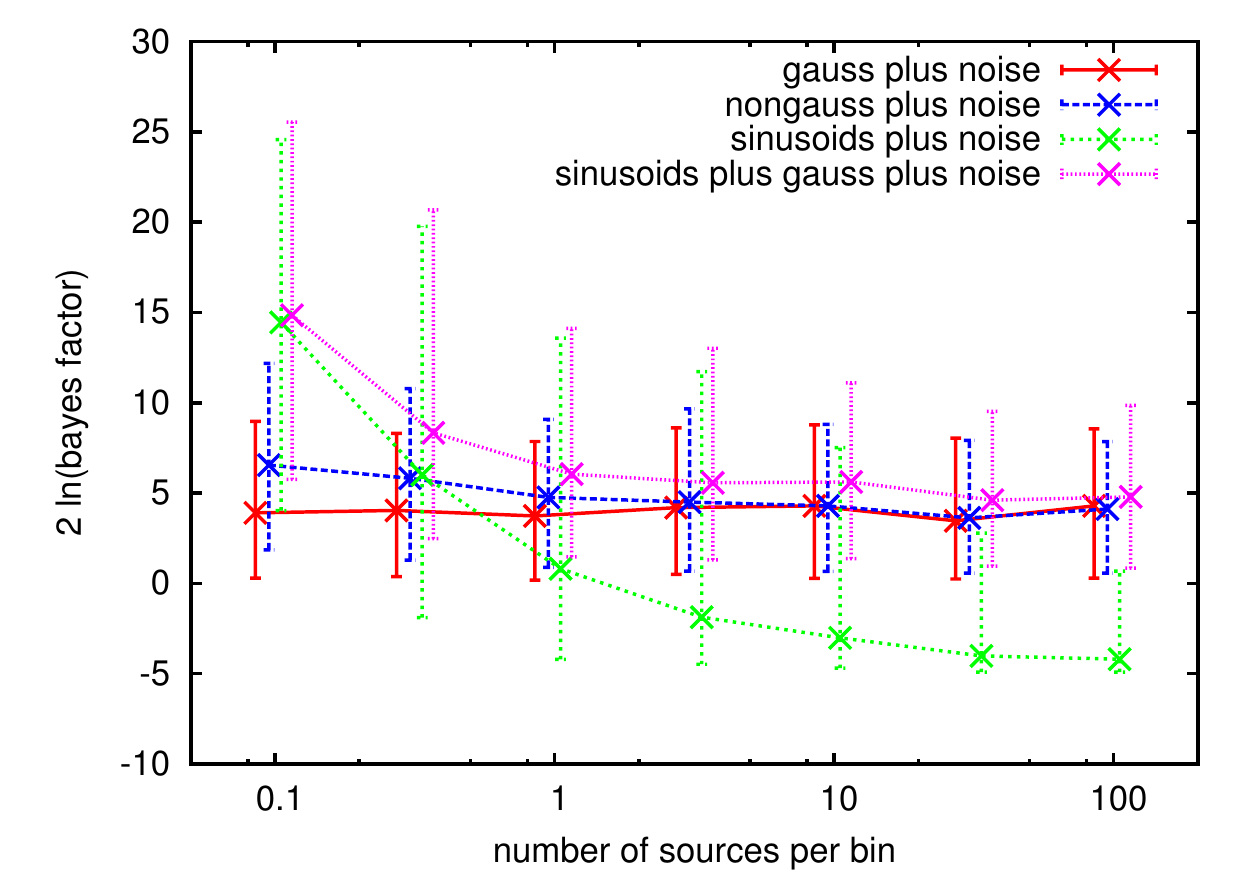}}
{\includegraphics[width=.49\textwidth]{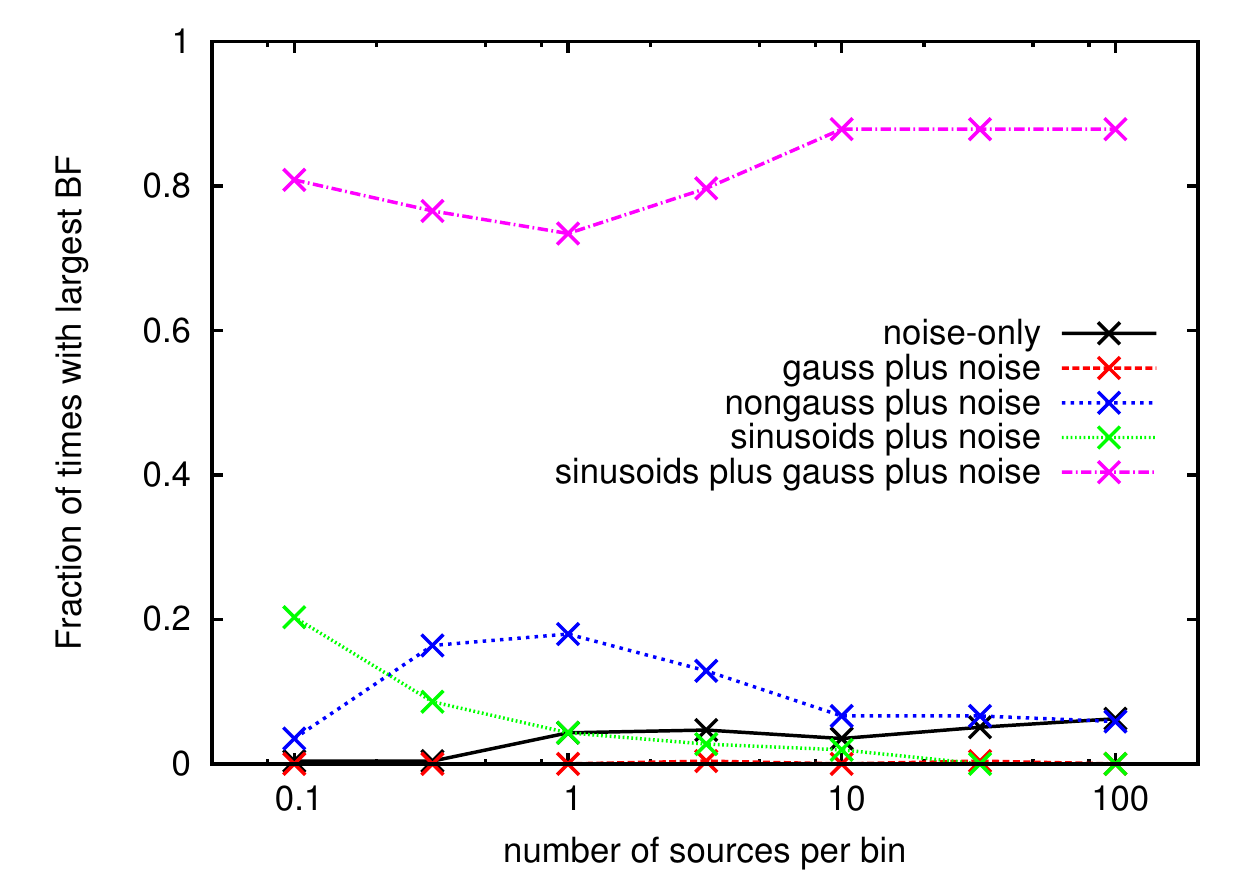}}
\caption{Upper panel: Bayes factor 80\% quantile intervals for the
  four different signal$+$noise models relative to the noise-only
  model as a function of the number of sources per bin.  The total
  number of bins was set to 32 for these simulations, and the
  astrophysical sources were drawn from source model 2.  The
  SNR-per-bin was fixed at 2, with different realizations of noise
  used for the different simulations.  Lower panel: Fraction of time
  that the different models had the largest Bayes factor for the
  different simulations.}
\label{f:SdenNoiseSourceModel2}
\end{figure}

Figure~\ref{f:SdenNoiseSourceModel2} extends the study of the
dependence on the source density to include the full set of signal
models, and in addition to showing the fraction of time that each
model is favored, also shows the Bayes factor quantile intervals for
the four signal models.  The total number of bins was set to 32 for
these simulations, which included simulated noise in addition to the
simulated astrophysical signals from source model 2.  For these
simulation the SNR-per-bin was fixed at 2, with different realizations
of the noise used for the different simulations.  Note that for low
source densities, the models that include deterministic sinusoid
signals are the preferred models.  The effectiveness of models having
Gaussian or non-Gaussian stochastic signal components improve as the
source density increases.  As expected, the hybrid model performs best
for all source densities.

%Figure~\ref{f:SdenNoiseSourceModel01} contains similar plots, 
%but for the other two astrophysical source models;
%model 0 has sources uniformly distributed in space out to some
%cutoff radius, and
%model 1 has sources are Gaussian distributed.
%The Bayes factors corresponding to these two astrophysical source
%models are typically smaller than those for the `Sesana' astrophysical model 
%shown in Fig.~\ref{f:SdenNoiseSourceModel2}, especially for 
%the sinusoids+noise and the sinusoids+gaussian+noise signal$+$noise 
%models at low source densities, where the brighter individual sinusoids
%stand out.
%
%\begin{figure*}[htbp]
%\begin{center}
%\subfigure[]{\includegraphics[width=.49\textwidth]{errorbars_SdenNoiseSourceModel0}}
%\subfigure[]{\includegraphics[width=.49\textwidth]{percentWins_SdenNoiseSourceModel0}}
%\subfigure[]{\includegraphics[width=.49\textwidth]{errorbars_SdenNoiseSourceModel1}}
%\subfigure[]{\includegraphics[width=.49\textwidth]{percentWins_SdenNoiseSourceModel1}}
%\caption{Same as Fig.~\ref{f:SdenNoiseSourceModel2}
%but for the other two astrophysical source models.
%Panels (a) and (b) are for source model 0, 
%where the sources are uniformly distributed in space out 
%to some cutoff radius.
%Panels (c) and (d) are for source model 1,
%where the sources are Gaussian distributed.}
%\label{f:SdenNoiseSourceModel01}
%\end{center}
%\end{figure*}
%

%
\begin{figure}[htbp]
{\includegraphics[width=.49\textwidth]{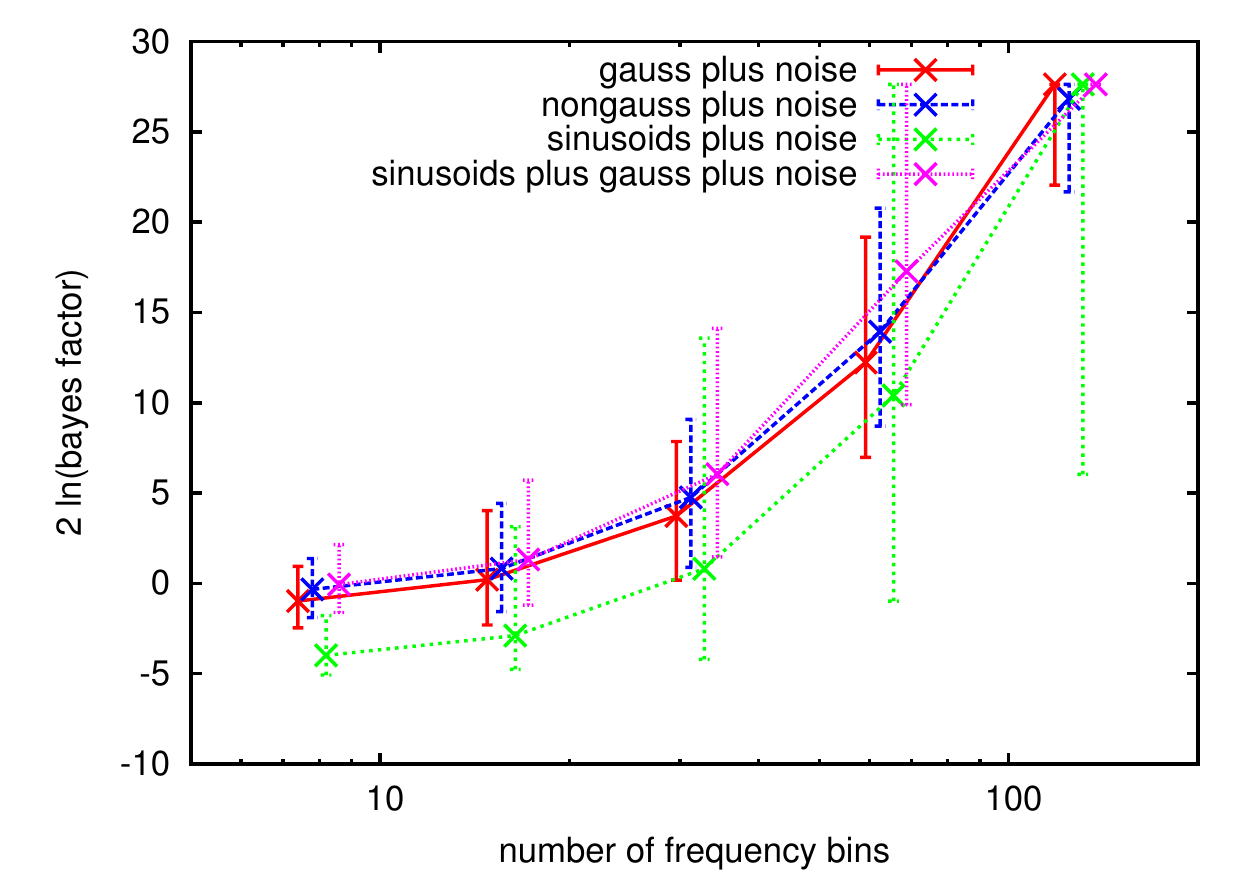}}
]{\includegraphics[width=.49\textwidth]{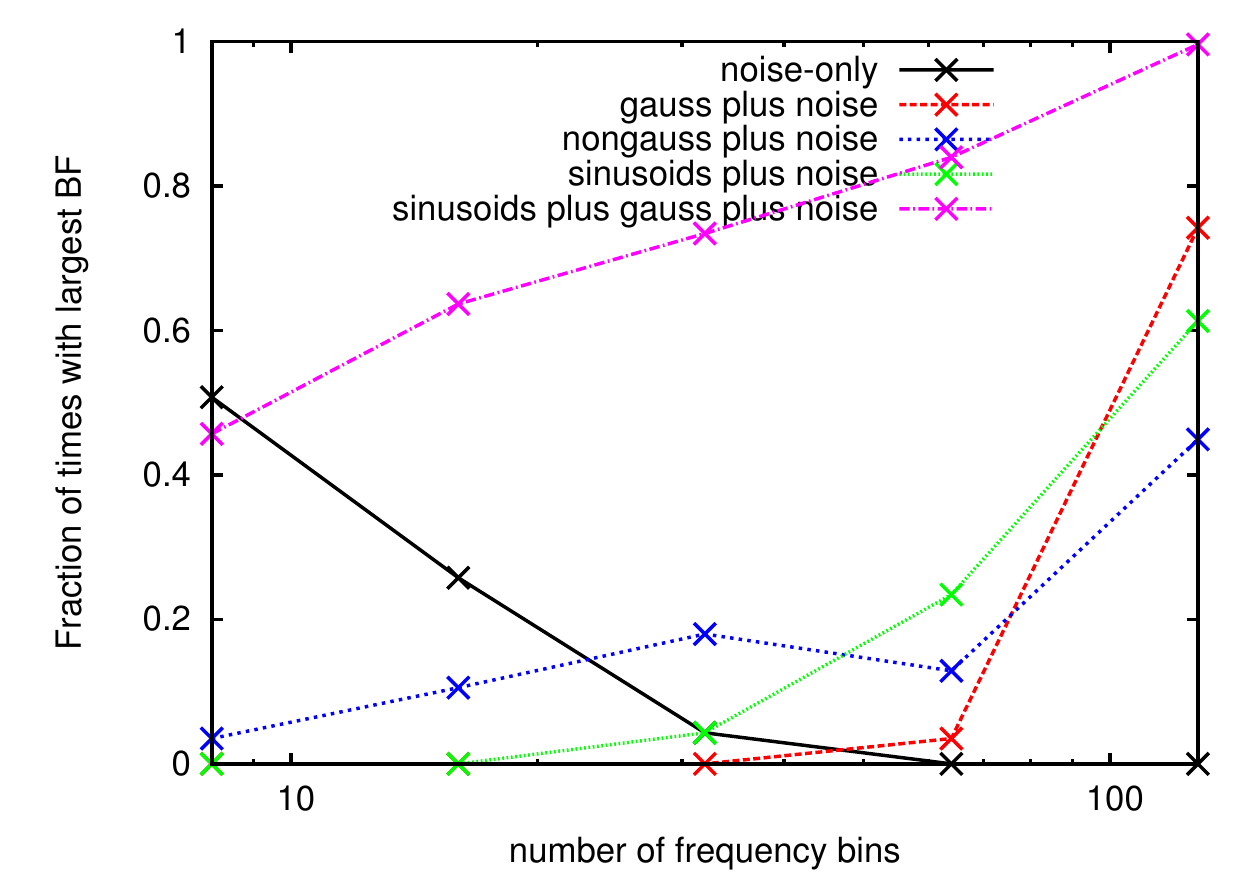}}
\caption{Upper panel: Bayes factor 80\% quantile intervals for the
  four different signal$+$noise models relative to the noise-only
  model as a function of the total number of frequency bins.  The
  source density was set to 1/bin for all the simulations, and the
  astrophysical sources were drawn from source model 2.  The
  SNR-per-bin was fixed at 2, with different realizations of noise
  used for the different simulations.  Lower panel: Fraction of time
  that the different models had the largest Bayes factor for the
  different simulations.}
\label{f:errorbars_NbinsNoise}
\end{figure}

Figure~\ref{f:errorbars_NbinsNoise} shows how the model selection
results are affected by the number of frequency bins, keeping the
source density fixed at one-per-bin and the SNR-per-bin fixed at 2. As
the number of frequency bins increases, the chances of having one or
two loud sources dominate the total signal increases, and
consequently, the deterministic multi-sinudoid model and the
two-component Gaussian stochastic models are more likely to outperform
the Gaussian model.

\begin{figure}[htbp]
{\includegraphics[width=.49\textwidth]{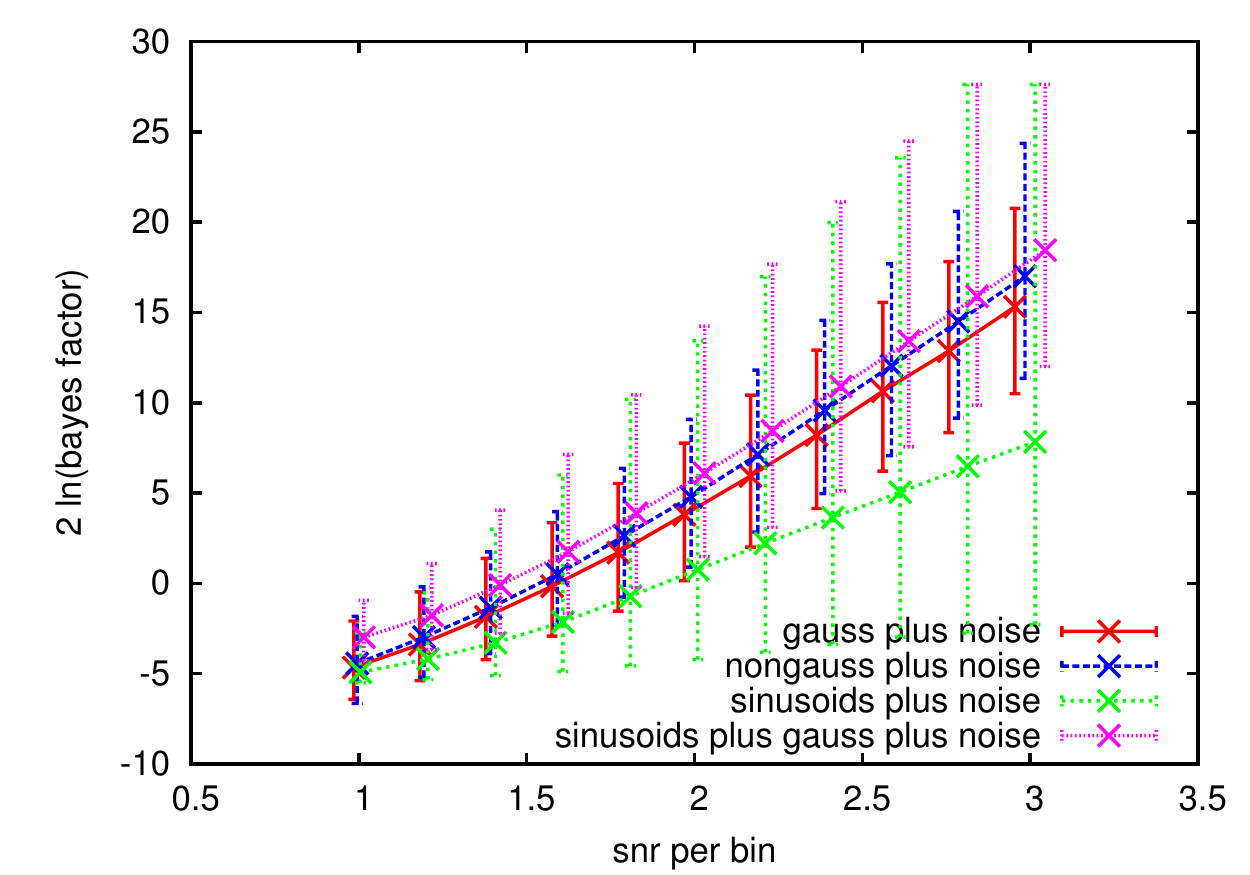}}
{\includegraphics[width=.49\textwidth]{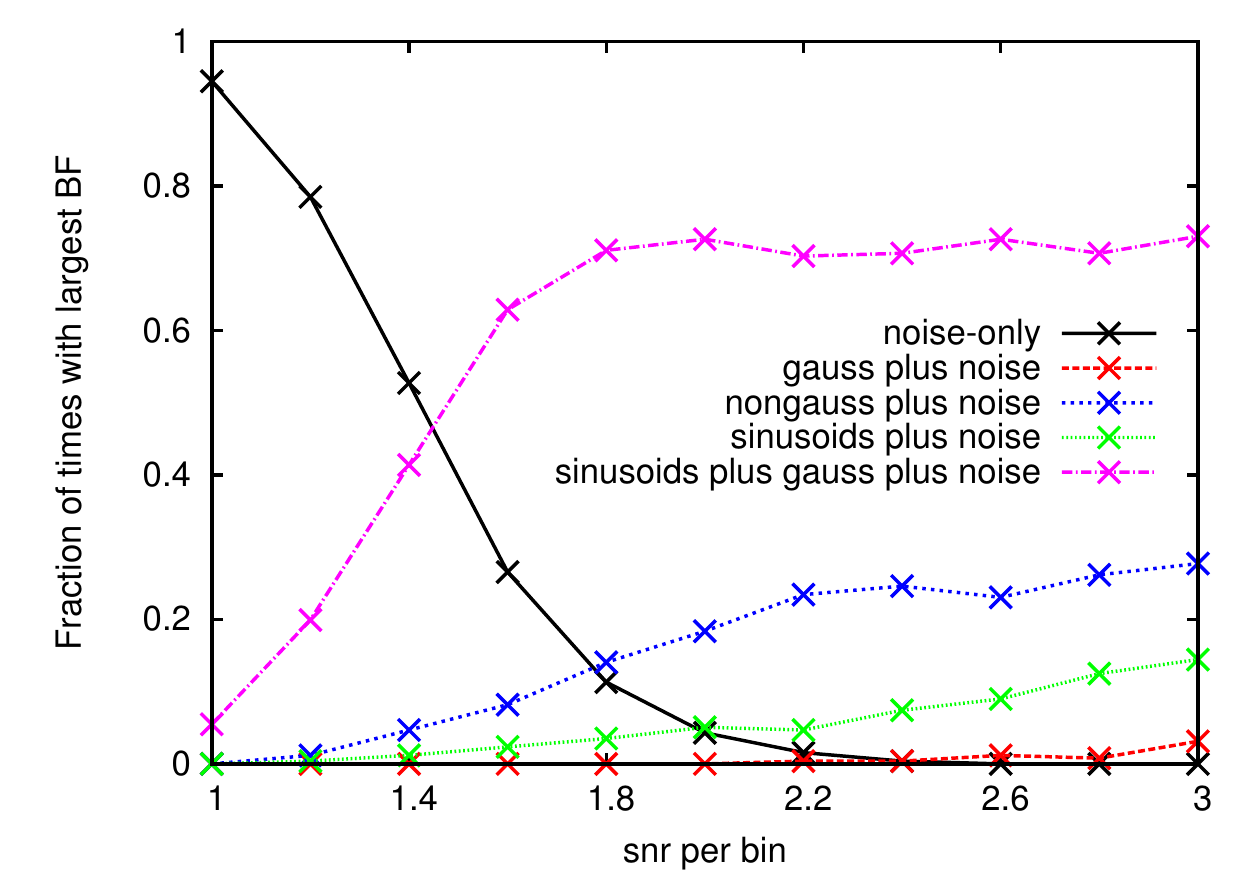}}
\caption{Upper panel: Bayes factor 80\% quantile intervals for the
  four different signal$+$noise models relative to the noise-only
  model as a function of SNR-per-bin.  The total number of bins was
  set to 32 and the source density to 1/bin for all the simulations.
  The astrophysical sources were drawn from source model 2.  Lower
  panel: Fraction of time that the different models had the largest
  Bayes factor for the different simulations.}
\label{f:errorbars_SNRNoise}
\end{figure}

Finally, Figure~\ref{f:errorbars_SNRNoise} shows how the model
selection is affected by the SNR-per-bin, keeping the source density
fixed at one-per-bin and the number of bins fixed at 32. The Bayes
factors for all the signal$+$noise models increase quadratically with
increasing SNR-per-bin, which is to be expected when compared to the
noise-only model. For sufficiently large SNR-per-bin (so that the
signal is detected by all the models), the relative performance of the
various signal models is independent of the SNR-per-bin.

%%%%%%%%%%%%%%%%%%%%%%%%%%%%%%%%%%%%%%%%%%%%
\section{Discussion}
\label{s:discussion}

We presented a Bayesian search for non-Gaussian gravitational-wave
backgrounds. We found that a gravitational wave signal comprised of
the sum of discrete sources drawn from some astrophysical population
may be best described as either deterministic, 
non-Gaussian stochastic, or Gaussian-stochastic 
depending on the number of sources, and the
size of the data set. In our studies the simulated data were produced 
by adding together multiple sinusoids with amplitudes drawn from one
of three astrophysical source distributions, to which was added an
independent white noise realization in each detector. While the
deterministic signal model ${\cal M}_3$, made up of a variable number
of sinusoids, is able to precisely match the simulated data, the
simpler Gaussian ${\cal M}_1$ and non-Gaussian ${\cal M}_2$ stochastic
signal models are often preferred. The general trend follows our
expectation that the signals appear increasingly stochastic and
Gaussian as the number of sources per frequency bin increases. We
found that departures from Gaussianity are more likely to be detected
in large data sets (in our case, for large numbers of frequency bins),
but that the ability to distinguish between the various signal models
was independent of the SNR (once the signal became detectable).  In
all cases, a hybrid model, ${\cal M}_4$, that combines variable
contributions from deterministic and stochastic signals that are
determined by the data, outperformed each of the single element models
most of the time. This finding may be of particular relevance to the
detection of low frequency gravitational waves by pulsar timing
arrays.

Although for simplicity we considered co-located and co-aligned
detectors and white power spectra, the method can easily be extended
to handle more realistic detector geometry (i.e., separated and
misaligned detectors) as well as colored spectra.  One can also
consider additional signal and noise models.  
\\ \\

%%%%%%%%%%%%%%%%%%%%%%%%%%%%%%%%%%%%%%%%%%%%%%%%%
\acknowledgments 
NJC acknowledges support from NSF Award PHY-1306702
and from the NANOGrav Physics Frontier Center, NSF PFC-1430284.  
JDR acknowledges support from NSF Awards PHY-1205585, 
CREST HRD-1242090 and the NANOGrav Physics Frontier Center, 
NSF PFC-1430284.  

%%%%%%%%%%%%%%%%%%%%%%%%%%%%%
\begin{appendix}
\section{Bayes factor calculation}
\label{a:bayesfactorapproxs}

Suppose we have two models that we would like to compare, 
denoted ${\cal M}_1$ and ${\cal M}_0$, with parameters 
$\vec\theta_1$ and $\vec\theta_0$, respectively.
As described in Sec.~\ref{s:modelselection}, the Bayes factor
${\cal B}_{10}(\mb s)$ for model ${\cal M}_1$ relative to model 
${\cal M}_0$ given observed data $\mb s$ is defined by
\begin{equation}
{\cal B}_{10}(\mb s) = 
\frac{p(\mb s|{\cal M}_1)}{p(\mb s|{\cal M}_0)}\,,
\end{equation}
where
\begin{equation}
p(\mb s|{\cal M}) = \int d\vec\theta\>
p(\mb s|\vec\theta, {\cal M})\pi(\vec\theta|{\cal M})
\label{e:modelint}
\end{equation}
for either model (i.e., ${\cal M}={\cal M}_0$ or ${\cal M}_1$).
The quantity $p(\mb s|{\cal M})$ is called the {\em evidence}
for  model ${\cal M}$.
The Bayes factor is the ratio of the evidences for the
two models; 
it equals the posterior odds ratio for the two models if 
they have equal a~priori probabilities.

Since analytic or direct calculations of the evidence integrals is
usually not possible, we need to estimate the Bayes factor
numerically.  In this appendix, we describe a few of the methods 
used for the study described in this paper.  
Readers interested in more details should see Sec.~II of 
\cite{Cornish-Littenberg:2007}.

%%%%%%%%%%%%%%%%%%%%%%%%%%%%%%%%%%%%%%%%%%%%%%%%%%%%
\subsection{Laplace approximation}
\label{s:laplace-approx}

If we assume that the data is informative, so that the likelihood
function is {\em peaked} relative to the prior probability
distribution, then we can use the {\em Laplace approximation} 
to estimate the evidence integral (\ref{e:modelint}):
\begin{equation}
p({\mb s}|{\cal M})
\simeq p({\mb s}|\vec\theta_{\rm ML},{\cal M})
\frac{\Delta V_{\cal M}}{V_{\cal M}}\,,
\end{equation}
where $\vec\theta_{\rm ML}\equiv\vec\theta_{\rm ML}(\mb s)$ 
maximizes the likelihood $p(\mb s|\vec\theta,{\cal M})$
with respect to variations of $\vec\theta$;
$\Delta V_{\cal M}$ is the characteristic width of the likehood 
function around its maximum; and $V_{\cal M}$ is the total 
parameter space volume for the model parameters.
The ratio $\Delta V_{\cal M}/V_{\cal M}$ can be thought of as 
an Occam's factor,
which penalizes a model if its parameter space volume is 
larger than needed to fit the data.
Doing this calculation for both ${\cal M}_0$ and ${\cal M}_1$,
and then taking the ratio of the two results, we find
\begin{align}
{\cal B}_{10}(\mb s) 
&\simeq
\frac{p({\mb s}|\vec\theta_{1,\rm ML},{\cal M}_1)}
{p({\mb s}|\vec\theta_{0,{\rm ML}},{\cal M}_0)}
\frac{\Delta V_1/V_1}{\Delta V_0/V_0}
\\
&=\Lambda_{\rm ML}(\mb s)
\frac{\Delta V_1/V_1}{\Delta V_0/V_0}\,,
\end{align}
where $\Lambda_{\rm ML}(\mb s)$ is the {\em maximum-likelihood ratio}.

As a very simple example, consider the case of $N$ samples of 
data $\mb s$, consisting of a unknown constant signal in additive
white Gaussian-stationary noise with known variance $\sigma$.
Let ${\cal M}_0$ denote the {\em noise-only} model with likelihood 
function
\begin{equation}
p(\mb s|{\cal M}_0) = 
\prod_{i=1}^N \frac{1}{\sqrt{2\pi \sigma^2}}
e^{-s_i^2/2\sigma^2}\,,
\end{equation}
and let ${\cal M}_1$ be the {\em signal$+$noise} model 
defined by the likelihood function 
\begin{equation}
p(\mb s|\theta, {\cal M}_1) 
= \prod_{i=1}^N \frac{1}{\sqrt{2\pi \sigma^2}}
e^{-(s_i-\theta)^2/2\sigma^2}\,,
\end{equation}
and prior $\pi(\theta)=1/\theta_{\rm max}$,
where $\theta\in [0,\theta_{\rm max}]$ is the unknown signal amplitude.
Then one can easily show that the maximum-likelihood parameter
value is the sample mean
\begin{equation}
\theta_{\rm ML}(\mb s) = \frac{1}{N}\sum_{i=1}^N s_i \equiv \bar s\,,
\end{equation}
and the Bayes factor for the signal$+$noise model ${\cal M}_1$ 
relative to the noise-only model ${\cal M}_0$ is:
\begin{equation}
{\cal B}_{10}(\mb s) \simeq 
\frac{\sigma/\sqrt{N}}{\theta_{\rm max}}
\exp\left[\frac{1}{2} \frac{\bar s^2}{\sigma^2/N}\right]\,.
\end{equation}
It has logarithm
\begin{equation}
2\ln {\cal B}_{10}(s) \simeq 
2\ln\left(\frac{\sigma/\sqrt{N}}{\theta_{\rm max}}\right)
+ \frac{\bar s^2}{\sigma^2/N}\,.
\label{e:BvsSNR}
\end{equation}
Since $\bar\sigma^2\equiv \sigma^2/N$ is the variance of 
the sample mean $\bar s$ (or, equivalently, it is the 
characteristic width of the likelihood function around its 
maximum value), we see that twice the log of the Bayes 
factor is effectively the squared SNR of the 
maximum-likelihood estimator $\theta_{\rm ML}({\mb s})$.
The first term on the right-hand side of Eq.~(\ref{e:BvsSNR})
is the Occam's penalty factor associated with size of the 
parameter space volume $\theta_{\rm max}$.  This term gets 
smaller as the number of data points $N$ increases since 
$\bar\sigma$ decreases with increasing $N$.

%%%%%%%%%%%%%%%%%%%%%%%%%%%%%%%%%%%%%%%%%%%
\subsection{Savage-Dicke density ratio}
\label{s:SDDR}

The Savage-Dicke density ratio can be defined whenever
model ${\cal M}_0$ is a subset of model ${\cal M}_1$,
and the prior probabilities factorize.
Both of these conditions hold, for example, if 
$\vec\theta_1=\{\vec\theta_0,\vec\theta_{\rm extra}\}$,
with
\begin{equation}
\pi(\vec\theta_1|{\cal M}_1)=
\pi(\vec\theta_0|{\cal M}_0)\pi(\vec\theta_{\rm extra}|{\cal M}_1)
\label{e:factor}
\end{equation}
and
\begin{equation}
p(\mb s|\vec\theta_0,{\cal M}_0)=
p(\mb s|\vec\theta_1,{\cal M}_1)
\big|_{\vec\theta_{\rm extra}=\vec\theta_{{\rm extra},0}}
\label{e:MtoM0}
\end{equation}
for some fixed set of parameter values $\vec\theta_{{\rm extra},0}$.
The Savage-Dicke density ratio $r_{10}(\mb s)$ is then defined as
\begin{equation}
r_{10}({\mb s})\equiv \frac{\pi(\vec\theta_{{\rm extra},0}|{\cal M}_1)}
{p(\vec\theta_{{\rm extra},0}|\mb s,{\cal M}_1)}\,,
\end{equation}
where $p(\vec\theta_{{\rm extra},0}|\mb s,{\cal M}_1)$ is the 
marginalized probability density function
\begin{equation}
p(\vec\theta_{\rm extra}|\mb s,{\cal M}_1)
=\int d\vec \theta_0\>
p(\vec\theta_0,\vec\theta_{\rm extra}|\mb s,{\cal M}_1)
\end{equation}
evaluated at $\vec\theta_{\rm extra}=\vec\theta_{{\rm extra},0}$.
Using Bayes' theorem in the form
\begin{equation}
p(\vec\theta_0,\vec\theta_{\rm extra}|\mb s,{\cal M}_1) 
= \frac{p(\mb s|\vec\theta_0,\vec\theta_{\rm extra},{\cal M}_1)
\pi(\vec\theta_0,\vec\theta_{\rm extra}|{\cal M}_1)}
{p(\mb s|{\cal M}_1)}
\end{equation}
and Eqs.~(\ref{e:factor}) and (\ref{e:MtoM0}), one can show that 
\begin{equation}
r_{10}(\mb s)={\cal B}_{10}(s)
\end{equation}
{\em exactly}.
The advantage of using the expression for the Savage-Dicke
density ratio to estimate the Bayes factor is that it only
requires exploration of the posterior distribution for model 
${\cal M}_1$.

%%%%%%%%%%%%%%%%%%%%%%%%%%%%%%%%%%%%%%%%%%
\subsection{Reversible jump MCMC}
\label{s:RJMCMC}

Reversible jump, or trans-dimensional MCMC algorithms, explore the
space of models in addition to the parameters of each model.  The
Bayes factor between two models ${\cal M}_0$ and ${\cal M}_1$ 
is simply estimated from the ratio of
the number of iterations that the chain spends in each model:
\begin{equation}
{\cal B}_{10}(\mb s) =\frac
{{\rm number\ of\ interations\ in\ model\ } {\cal M}_1}
{{\rm number\ of\ interations\ in\ model\ } {\cal M}_0}
\end{equation}
The accuracy of the estimate depends on the number of transitions
between the two models---the more transitions, the more accurate the
estimate. The problem with this simple approach is that it becomes
difficult to compute Bayes factors smaller than $10^{-3}$ or larger
than $10^3$, since the chains spend very little time in the
dis-favoured model, and hence the exploration of that model can fail
to converge to the stationary state. Ideally, we would like the chain
to spend an equal amount of time in each model, so that all models are
explored equally well (that is, assuming each model has a comparable
dimensionality; if the model dimensions are significantly different,
more time should be spent exploring in the higher-dimensional
model). 

To achieve good mixing within each model and between models,
we introduce an artificial prior weighting on the models that
compensates for the difference in the Bayes factors. For example, if
the Bayes factor between two models is 1000, we introduce a prior that
favors the low probability model by a factor of 1000, so the chains
spends an equal number iterations in each model~\cite{fakeprior}. Since the appropriate
weighting is not known in advance, an iterative scheme is used that
adjusts the artificial prior weighing on the models until balance is
achieved.  The true Bayes factors are then found from the iteration
ratio divided by the artificial prior odds ratio.

\end{appendix}

% use bibtex
\bibliography{references}

\end{document}